\documentclass[fleqn,usenatbib]{mnras}
\usepackage{newtxtext,newtxmath}
\usepackage[T1]{fontenc}
\usepackage{graphicx}	
\usepackage{amsmath}	
\usepackage{CJK}
\newcommand{\alf}{Alfv$\acute{\text{e}}$n } 

\newcommand\bb[1]{\mbox{\boldmath{$#1$}}}

\newcommand{\D}[2]{\frac{{\rm d} #2}{{\rm d} #1}}

\newcommand\grad{\bb{\nabla}}
\newcommand\bcdot{\,\bb{\cdot}\,}

\newcommand\eb{\hat{\bb{b}}}
\newcommand\ez{\hat{\bb{z}}}
\newcommand\rmd{{\rm d}}
\newcommand\imag{{\rm i}}
\usepackage[none]{hyphenat}

\defcitealias{Meyer1994}{MM94}
\defcitealias{Cho2022}{CN22}
\defcitealias{Bambic2024}{BQK24}

\begin{document}

\title[Conduction in accretion disc coronae]{Local models of two-temperature accretion disc coronae. II. Ion thermal conduction and the absence of disc evaporation}

\author[C.~J.~Bambic, E.~Quataert, M.~W.~Kunz, \& Y.~Jiang]  
    {\parbox[]{7.in}{Christopher~J.~Bambic$^{1}$\thanks{E-mail: 
          cbambic@princeton.edu}, Eliot~Quataert$^1$, Matthew~W.~Kunz$^{1,2}$ and
          \begin{CJK*}{UTF8}{gbsn}
          Yan-Fei Jiang (姜燕飞)$^{3}$\\
          \end{CJK*}
    \footnotesize 
    $^1$ Department of Astrophysical Sciences, Peyton Hall, Princeton University, Princeton, NJ 08544, USA \\
    $^2$ Princeton Plasma Physics Laboratory, PO Box 451, Princeton, NJ 08543, USA \\ 
    $^3$ Center for Computational Astrophysics, Flatiron Institute, New York, NY 10010, USA \\
  }
}

\maketitle

\begin{abstract}
We use local stratified shearing-box simulations with magnetic field-aligned thermal conduction to study an idealized model of the coupling between a cold, radiatively efficient accretion disc, and an overlying, hot, two-temperature corona. Evaporation of a cold disc by conduction from the hot corona has been proposed as a means of mediating the soft-to-hard state transitions observed in X-ray binary systems. We model the coronal plasma in our local disc patch as an MHD fluid subject to both free-streaming ion conduction and a parameterized cooling function that captures the collisional transfer of energy from hot ions to colder, rapidly cooling leptons. In all of our models, independent of the initial net vertical magnetic flux (NF) threading the disc, we find no evidence of disc evaporation. The ion heat flux into the disc is radiated away before conduction can heat the disc's surface layers. When an initial NF is present, steady-state temperature, density, and outflow velocities in our model coronae are unaffected by conduction. Instead of facilitating disc evaporation, thermal conduction is more likely to feed the disc with plasma condensing out of the corona, particularly in flows without NF. Our work indicates that uncertainties in the amount of NF threading the disc hold far greater influence over whether or not the disc will evaporate into a radiatively inefficient accretion flow compared to thermal conduction. We speculate that a change in net flux mediates disc truncation/evaporation.

\end{abstract}


\section{Introduction}
\label{sec:intro}

Thermal conduction from hot to cold gas can act to evaporate the cold medium, driving an evaporative flow \citep{Cowie1977, McKee1977_evaporation, Doroshkevich1981, Giuliani1984, Draine1984, Balbus1986_conduction}. In the absence of cooling and gravity, the thermal energy flux provided by conduction into the cold medium must be balanced by the enthalpy flux of the ensuing evaporative outflow \citep*{Draine2011}. These ideas, dating back to work on the solar corona by Pikel'ner~(1948, 1950), have found widespread application to a diverse range of astrophysical phenomena, from the destruction of cold clouds embedded in hot supernova remnants to accretion discs around white dwarfs and black holes.

In this paper, our focus is on accretion discs in X-ray binary systems (XRBs) and active galactic nuclei (AGN). XRBs exhibit a variety of spectral states, most notably, the `high/soft' and `low/hard' states, where soft and hard refer to the spectral `hardness' of power-law emission in the X-ray band (see \cite*{Remillard2006} and \cite*{Done2007} for reviews). 

Transitions from the soft to the hard state are often described via the disc truncation model of \cite*{Esin1997}. In this model, the luminous soft state is associated with a standard thin accretion disc \citep*{Shakura1973} extending down to the innermost stable circular orbit (ISCO) of a stellar mass black hole, while the low-luminosity hard state is associated with a truncated disc feeding a hot, optically thin, radiatively inefficient accretion flow \citep[RIAF;][]{Ichimaru1977, Rees1982, Narayan1994}. A hot (electron temperature $T_e \sim 10^9$~K) corona overlying the thin disc in the soft state, as would be expected in the observationally motivated two-phase model of a corona `sandwiching' an accretion disc \citep{Haardt1991}, provides a reservoir of thermal energy that can be tapped to evaporate the thin disc and form a RIAF \citep{Liu1999}. The evaporation itself occurs through thermal conduction from the corona into the disc \citep*{Meyer2000}, an idea first invoked to describe salient features in the phenomenology of dwarf novae in accreting white dwarf stars, i.e., cataclysmic variable stars \citep[hereafter, \citetalias{Meyer1994}]{Meyer1994}. \textit{If} the heat flux from the corona into the disc is large enough to overcome cooling, disc evaporation and truncation can ensue. Thus, thermal conduction offers a physical mechanism for mediating disc truncation, and subsequently, state transitions, in XRBs.

Conductive evaporation has been extended to XRB and AGN accretion flows by \cite{MeyerHofmeister1999} and \cite{Liu1999}; however, even with the recognition that coronae in accretion flows around black holes should be two-temperature, with the ion temperature much larger than that of the electrons \citep*{DiMatteo1997}, the vast majority of works on XRBs and AGN have restricted their attention to \textit{electron} thermal conduction \citep{Rozanska2000, Liu2002}, rather than considering the heat flux carried by ions. This assumption is justified in the colder transition region just above the disc; however, ion thermal conduction is certainly critical in the hot, weakly collisional corona \citep*{Spruit2002, Dullemond2005}, and even the temperature profiles of \cite{Liu2002} imply that ion thermal conduction would dominate over that of the electrons. 

Electron thermal conduction has been shown to be inefficient at inducing evaporation within a few tens of gravitational radii $R_g$ of a central accreting compact object, where $R_g \equiv G M/c^2$ and $M$ is the mass of the compact object. At distances of ${\gtrsim} 10^3 R_g$, evaporation is far more efficient, and the mass-outflow rate provided through conductive evaporation can outpace accretion. Equivalently, since the surface of a white dwarf with a mass $M \approx 1.2 {\rm M}_{\odot}$ is at ${\approx}10^3 R_g$, conduction can induce evaporation in the inner regions of white dwarf discs. \citet[][hereafter \citetalias{Cho2022}]{Cho2022} recently revisted this problem, analyzing the interplay of electron thermal conduction and bremmstrahlung cooling in accretion flows around black holes. Similar to previous works, these authors find that evaporation is inefficient at small radii near the black hole. Instead, conduction works to cool coronae, forming a condensing flow onto the disc. This process of conductive condensation could re-form a disc from a RIAF near the ISCO and influence the hard-to-soft transition in XRBs (\citealt{Liu2007}; \citealt*{Meyer2007}).

It is unclear whether a mechanism of disc truncation that operates at large radii alone can explain the observed properties of XRBs. Observations clearly indicate that the disc truncates at a few tens of $R_g$ in the hard state of Cygnus X-1, the best studied of the XRBs \citep{Zdziarski1999, Churazov2001_XRBs}; however, a number of other XRBs may possess a cold disc near the ISCO, even in the hard state \citep{Miller2006, Kara2019}. If discs are in fact truncated on scales of $\sim \rm{few} \times 10 \: R_g$, then a physical mechanism beyond electron thermal conduction is necessary to induce disc evaporation and truncation in XRB and AGN accretion flows. 

This paper explores the conductive coupling between a cold disc and hot corona to determine if evaporation can occur in the innermost regions of XRB and AGN accretion discs. Because our work is focused on the innermost few gravitational radii of these accretion flows, the thermodynamic regime explored here is that of a two-temperature ion--lepton plasma. In Paper I of this series, we demonstrated that, in this regime, the relevant cooling mechanism for the ions is Coulomb collisions with rapidly Compton-cooled leptons, and \textit{ion}, rather than electron, thermal conduction in the saturated, free-streaming limit dominates the transport of heat \citep*[hereafter, \citetalias{Bambic2024}]{Bambic2024}. These thermodynamic processes, distinct from those studied by \citetalias{Cho2022}, offer a possible means for truncating XRB and AGN discs close to the central object.

While conductive evaporation and disc truncation are the main motivations for this work, conduction may produce a variety of other observable effects in black hole accretion flows that can be explored by our models. \cite{Spruit2002} contend that the hard X-ray continuum emission characteristic of coronae results from the bombardment of electrons by MeV protons in a two-temperature RIAF. Ion conduction is a natural way to channel these hot ions into the X-ray emitting electrons. Conduction may affect the rate of energy release in coronae. \cite{Goodman2008} argue that the rate of magnetic reconnection may be regulated by conductive evaporation, which acts to modify the plasma density in the corona. Similar processes may occur on the Sun \citep{Uzdensky2007}, where nearly half of the thermal energy released in a flare leaves the reconnecting X-point via conduction. Finally, there have been claims that a truncated disc can still produce a detectable iron line at the ISCO \citep*{Tanaka1995, Reis2010} if colder, denser clumps can form within a RIAF \citep{Liska2022}. However, just as conduction may act to destroy cold clouds within supernova remnants \citep{Cowie1977, McKee1977_evaporation}, so too may conduction erase this signal.  

This paper is organized as follows. In \S\ref{sec:theory}, we introduce models for two-temperature cooling and free-streaming thermal conduction relevant for the innermost regions of XRB and AGN accretion flows. We then apply these models to a suite of vertically stratified magnetohydrodynamic (MHD) shearing-box simulations \citep{Stone1996} that, for the first time, include field-aligned free-streaming ion thermal conduction. The set-up of these simulations is described in \S\ref{sec:methods}, while the results are presented in \S\ref{sec:results}. In \S\ref{sec:toy_model}, we explore a series of toy models in one dimension to better understand our results, namely, that the inclusion of saturated ion thermal conduction cannot evaporate our model thin accretion discs. We discuss the implications of our results to state transitions in \S\ref{sec:discussion}, and we summarize and conclude in \S\ref{sec:conclusion}.

\section{Conduction in a Two-temperature Corona} \label{sec:theory}

We begin by introducing models for free-streaming ion thermal conduction and Coulomb collisional cooling relevant for the thermodynamic properties of plasma in the innermost regions of XRB and AGN accretion flows. Using these models, we derive an order-of-magnitude estimate for the conditions under which conductive heating can win out over two-temperature cooling to induce evaporation of an accretion disc. These estimates motivate three-dimensional MHD simulations that can evaluate if conductive evaporation can ever be realized.

Ions exchange energy with leptons on the temperature equilibration timescale, $t_{\rm eq}$, defined in \citetalias{Bambic2024} as
\begin{equation} \label{eq:temp_equilibration}
\begin{split}
    t_{\rm eq} &= \frac{1}{2} \frac{m_i}{m_e} \frac{3 \sqrt{m_e} (k_{\rm B} T_e)^{3/2}}{4 \sqrt{2\pi} n_e e^4 \ln{\Lambda_e}} \\
    &\approx 
    5 \times 10^{-3} \: \left(\frac{\Theta_e}{0.2} \right)^{3/2} 
    \left( \frac{n_e}{10^{17} \: {\rm cm}^{-3}} \right)^{-1} \left( \frac{\ln{\Lambda_e}}{23} \right)^{-1} \: {\rm s},
\end{split}
\end{equation}
where the lepton temperature $\Theta_e \equiv k_{\rm B} T_e/m_e c^2$ is defined in terms of the electron rest mass $m_e c^2$, the lepton number density is $n_e$, and we have introduced the electron Coulomb logarithm $\Lambda_e$. A simple estimate based on observational implications that the optical depth to electron scattering $\tau_{\rm es}$ is ${\sim}\mathcal{O}(1)$ in coronae implies that the lepton number density is $n_e \approx 10^{17} (M_{\rm BH}/ 10~{\rm M}_{\odot})^{-1} \: {\rm cm}^{-3}$.

We describe the cooling of ions through a simple, optically thin cooling function $\Lambda$ of the form 
\begin{equation}\label{eq:cooling}
    \D{t}{\mathcal{E}_{\rm int}} = - \Lambda = - \frac{\mathcal{E}_{\rm int}}{t_{\rm eq}},
\end{equation}
where $\mathcal{E}_{\rm int} = (3/2) n_i k_{\rm B} T$ is the internal energy of the plasma, $n_i$ is the ion number density, and $T$ is the ion temperature, with $T~\gg~T_e$ \citep*{DiMatteo1997}. Observations constrain the electron temperature and the optical depth in the corona, such that the product of the equilibration time and the Keplerian orbital frequency, $\Omega \approx 6.4 \times 10^2 \: \left( M/10~{\rm M}_{\odot} \right)^{-1} \left( R/ 10~R_g \right)^{-3/2} \: {\rm s}^{-1}$, is given by
\begin{equation} \label{eq:Omega_teq_obs}
    \Omega t_{\rm eq} = 7 \: \chi \: \tau_{\rm es}^{-1} \left( \frac{\Theta_e}{0.2} \right)^{3/2} \left( \frac{\ln{ \Lambda_e }}{23} \right)^{-1} \left( \frac{R}{10 \: R_{\rm g}} \right)^{-1/2}.
\end{equation}
Here, $R$ is the radial size of the corona, while $\chi$ is an $\mathcal{O}(1)$ (possibly $\mathcal{O}(H/R)$, where $H$ is the disc scale height) factor set by the unknown geometry of the corona. Based on the optical depths implied by observations, Coulomb collisions should be weak enough to allow ions to heat up to an order unity fraction of the virial temperature $k_{\rm B} T_{\rm virial} \equiv (1/3) m_i c^2 (R/R_{\rm g})^{-1}$, where $m_i$ is the proton mass. At $10\: R_{\rm g}$, this ion temperature is $T_i~\approx 31~{\rm MeV} \approx 4 \times 10^{11} \: {\rm K}$. 

Magnetic-field-aligned ion thermal conduction in the free-streaming limit can be captured via a heat flux of the form \citep*{Malone1975}
\begin{equation} \label{eq:F_free}
    \boldsymbol{F}_{\rm con}^{\rm free} = 0.6 \rho c_{s}^3 \: {\rm sgn} \left( (\ez \bcdot \eb) L_T \right) \eb.
\end{equation}
Here, the mass density of the coronal ions is $\rho$, the speed of sound $c_s$ is defined via $c_s^2 = k_{\rm B} T/m_i$, the thermal gradient length scale is defined  as $L_T \equiv -(\mathrm{d} \ln{T}/ \mathrm{d} z )^{-1}$, the unit vector aligned with the magnetic field $\boldsymbol{B}$ is $\eb \equiv \boldsymbol{B}/|\boldsymbol{B}|$, and we work in a local coordinate system in our disc patch such that $z$ is the vertical coordinate with unit direction $\ez$.

The free-streaming heat flux is appropriate when ion collisions are sufficiently infrequent that the \cite{Spitzer1962} collisional heat flux saturates at the free-streaming limit. We can write the Spitzer heat flux as 
\begin{equation} \label{eq:F_Spitzer}
    \boldsymbol{F}_{\rm con}^{\rm Spitzer} = \frac{3.9 P}{L_T} \frac{k_{\rm B} T}{m_i} t_{ii} \eb,
\end{equation}
where $P = n_i k_{\rm B} T$ is the ion pressure and the ion--ion collision time $t_{ii}$ is longer than the equilibration time \eqref{eq:temp_equilibration} by a factor of ${\sim}(T_i/T_e)^{3/2} (m_e/m_i)^{1/2}$. For our estimate of $T_i \simeq T_{\rm virial}$, we find that $t_{\rm ii}/ t_{\rm eq} \approx 130$. The Spitzer heat flux saturates for any thermal gradient length scale satisfying
\begin{eqnarray}
    \frac{|L_{T}|}{H_{z, {\rm c}}} &\lesssim& 5 \left( \frac{T_i}{T_e} \right)^{3/2} \left( \frac{m_e}{m_i} \right)^{1/2} \Omega t_{\rm eq} \\
    &=& 4 \times 10^3 \: \frac{\chi}{\tau_{\rm es}} \left( \frac{\Theta_e}{0.2} \right)^{3/2} \left( \frac{\ln{ \Lambda_e }}{23} \right)^{-1} \left( \frac{R}{10 \: R_{\rm g}} \right)^{-1/2},
\end{eqnarray}
where $H_{z, {\rm c}} \equiv \sqrt{2} c_s/ \Omega$ is the vertical scale height of the corona. Thus, even small temperature gradients with large $|L_T|$ will still saturate the heat flux at the free-streaming value. Similarly, the large ion-to-lepton temperature ratio implies that, unlike for the scenarios studied in \citetalias{Meyer1994}, \cite{Liu2002}, \cite{Liu2007}, \cite*{Meyer2007}, and \citetalias{Cho2022}, electron thermal conduction is negligible. Even at the saturated limit, ion thermal conduction is larger than that of electrons by the same factor of $(T_i/T_e)^{3/2} (m_e/m_i)^{1/2} \approx 130$, such that ion conduction carries the bulk of the conductive heat flux within a two-temperature corona. 

For conduction to evaporate an accretion disc, the energy released in the disc must overcome two-temperature cooling and heat the disc, driving a substantial enthalpy flux into the corona. Conduction overcomes two-temperature cooling when 
\begin{equation} \label{eq:evaporation_criterion}
\begin{split}
    &\frac{\left|\grad \bcdot \boldsymbol{F}_{\rm con}^{\rm free}\right|}{\Lambda} \gtrsim 0.3 f_s \left( \frac{L_T}{H_{z, {\rm c}}} \right)^{-1} \Omega t_{\rm eq} \gtrsim 1, \\
    &\implies \frac{|L_T|}{H_{z,{\rm c}}} \lesssim 0.3 f_s \Omega t_{\rm eq}.
\end{split}
\end{equation}
Here, we have introduced a geometric suppression factor $f_s$ to account for the fact that conduction across the magnetic field is strongly suppressed relative to conduction along the magnetic field (see Paper I). In principle, this condition could be satisfied in the corona where $\Omega t_{\rm eq} \gtrsim 1$, so long as the suppression factor is not ${\ll}1$. However, in the disc's surface layers, where densities are large enough that $\Omega t_{\rm eq} \ll 1$, Coulomb collisions will rapidly cool the ions before evaporation can occur. To determine the temperature gradient length scales $L_T$, suppression factors $f_s$, and Coulomb cooling timescales $\Omega t_{\rm eq}$ in an accretion disc, we turn to numerical simulations. 

\section{Methods} \label{sec:methods}

The simulation set-up, initial conditions, and boundary conditions imposed on the MHD variables are identical to those presented in \citetalias{Bambic2024}. To implement conduction, we must include the conductive heat flux $\boldsymbol{F}_{\rm con}$ in the MHD energy equation and evaluate an equation for the evolution of $\boldsymbol{F}_{\rm con}$. Here, we describe a two-moment method, as presented in \cite{Jiang_Oh2018} and originally applied to the problems of radiation transport and cosmic-ray MHD, which we use to evolve the free-streaming heat flux in our simulations. We can then solve the full system of equations for MHD, thermal conduction, and cooling within a local approximation for an accretion disc, the stratified shearing box \citep*{Hawley1995}, using the \textit{Athena++} MHD code \citep{Stone2020}. The shearing box `zooms in' on a local patch of plasma orbiting a black hole at a radius $R_0$ and the Keplerian angular velocity $\boldsymbol{\Omega} = \Omega \ez$ through mapping the global coordinates $(r,\varphi,z)$ to Cartesian coordinates $x = r - R_0$, $y = R_0 (\varphi - \Omega t)$, and $z = z$, where the simulation evolves with time $t$. 

\subsection{The Coulomb cooling function} \label{sec:cooling_function}

Rather than evolve separate energy equations for the ions and leptons, we treat the ions as a single MHD fluid subject to cooling via Coulomb collisions. Our goal is to model properly the approximately virialized ions in the surface layers of the disc while keeping the mid-plane of the simulation at a fixed temperature and thus, scale height. We have already introduced a cooling function in Equation~\ref{eq:cooling}. By ignoring the Coulomb logarithm, which depends only weakly on $n_e$ and $T_e$, we arrive at a cooling function that can be implemented in our simulations,
\begin{equation} \label{eq:cooling_function}
    \D{t}{\mathcal{E}_{\rm int}} = -\Lambda (\rho, T) = -2 {\mathcal{A}} \left( \frac{\rho}{\rho_0} \right)^2 \left( \frac{T}{T_0} \right) \: \rho_0 T_0 \Omega \:\:{\rm for}\:\: T > T_0.
\end{equation}
Here, $k_{\rm B} = m_i = 1$, density $\rho$ and temperature $T$ are measured relative to their mid-plane values, $\rho_0$ and $T_0$ respectively, and the uncertain lepton physics is absorbed into a constant free parameter $\mathcal{A}$, which we vary in our model (\citetalias{Bambic2024}). For weak Coulomb coupling, i.e., $\mathcal{A} \sim 10-10^2$, this form for the cooling function ensures that the cooling time will be very short in the bulk of the disc, where $\rho \simeq \rho_0$; however, the cooling time can become long relative to the orbital timescale in the diffuse surface layers of the disc where $\rho \ll \rho_0$, consistent with the properties of ions in the corona. 

\subsection{Equations solved} \label{sec:equations}

Traditional methods for thermal conduction suppose a closure relation for $\boldsymbol{F}_{\rm con}$ and treat $-\grad \bcdot \boldsymbol{F}_{\rm con}$ as a source term in the total energy equation. In the free-streaming regime relevant to ion conduction in accretion disc coronae, standard explicit methods for computing this source term introduce spurious, large-amplitude  oscillations in the heat flux, which can only be controlled through a rather stringent condition on the timestep, $\Delta t \sim (\Delta x)^3$ \citep*{Sharma2010}. Such an onerous constraint is computationally prohibitive.

The two-moment method that we employ \citep{Jiang_Oh2018, Tan2021} circumvents this timestep constraint by introducing a maximum signal velocity $v_{\rm max}$, which controls the timestep via the standard Courant condition.  Conduction acts to modify the internal energy of the MHD fluid through
\begin{equation} \label{eq:first_moment}
    \frac{\partial \mathcal{E}_{\rm int}}{\partial t} = - \grad \bcdot \boldsymbol{F}_{\rm con} .
\end{equation}
This is the first of the moment equations. Assuming a closure for the equilibrium, time-steady heat flux along field lines, $\boldsymbol{F}_{\rm con} = - \kappa \eb\eb \bcdot \grad T $ \citep{Braginskii1965}, where $\kappa$ is a conductivity that depends on local plasma conditions, the heat flux can be evaluated via solution of the second moment equation,
\begin{equation} \label{eq:second_moment}
    \frac{1}{v_{\rm max}^2} \frac{T}{\mathcal{E}_{\rm int}} \frac{\partial \boldsymbol{F}_{\rm con}}{\partial t} +  \eb \eb \bcdot \grad T = -\frac{1}{\kappa} \boldsymbol{F}_{\rm con}.
\end{equation}
This moment equation reduces to the desired form of the heat flux, $\boldsymbol{F}_{\rm con} = - \kappa \eb \bcdot \grad T \eb$, in the limit
\begin{equation} \label{eq:vmax_condition}
    \left( \frac{c_s}{v_{\rm max}} \right)^2 \frac{\kappa}{\mathcal{E}_{\rm int}} \frac{\partial}{\partial t} \ln{ \left( \boldsymbol{F}_{\rm con} \right) } \ll 1.
\end{equation}
Crucially, Equation~\eqref{eq:second_moment} describes the saturated, free-streaming limit of thermal conduction when $\kappa$ is chosen such that
\begin{equation} \label{eq:kappa}
    \kappa = {\rm max} \left( \frac{0.6 \rho c_s^3}{|\eb \bcdot \grad T|}, \: \kappa_{\rm min} \right),
\end{equation}
where $\kappa_{\rm min} = 10^{-10} \: \rho_0 H_z^2 \Omega$ is a negligibly small conductivity included for numerical stability. All simulations with free-streaming thermal conduction use this expression for $\kappa$, and the two-moment equations (\ref{eq:first_moment} and \ref{eq:second_moment}) are solved to evaluate the heat flux. We demonstrate the accuracy of this method in Appendix~\ref{sec:sound_waves}.

\subsection{Initial Conditions and Boundary Conditions} \label{sec:ICs_BCs}

We adopt units in which the Keplerian orbital frequency in our local disc patch is $\Omega=1$, the mid-plane density $\rho_0 = 1$, and the mid-plane temperature $T_0 = 1/2$. The scale height of the disc at the mid-plane, $H_z$, is determined through vertical force balance:
\begin{equation} \label{eq:Hz}
    H_z^2 = \frac{2 c_{\rm s0}^2}{\Omega^2} = 1,
\end{equation}
while hydrostatic equilibrium with the local potential $\Phi$ determines the initial density distribution. Pressure is determined by assuming the atmosphere is initially at a single temperature $T_0$. Time is measured relative to the orbital timescale $t_{\rm orb} = 2 \pi/ \Omega = 2\pi$. 

The conductive heat flux is initially set to $\boldsymbol{F}_{\rm con} = 0$ throughout the domain, corresponding to the fact that we have assumed an initially uniform-temperature atmosphere. To seed the magnetorotational instability \citep[MRI;][]{Balbus1991} in our simulations, we initialize the domain in the region $|z| \leq 0.5 H_z$ with randomly distributed, spatially uncorrelated velocity and adiabatic pressure fluctuations with maximum amplitude $\delta \boldsymbol{v} = 5 \times 10^{-3} c_{\rm s}$ and $\delta P/P = 0.025$ \citep{Hawley1995}. These small perturbations introduce a small heat flux initially in the domain; however, cooling and conduction rapidly smooth out these initial temperature fluctuations, and our results are independent of the initial values for the heat flux.

While the boundary conditions of the density $\rho$, pressure $P$, velocities $\boldsymbol{u}$, and magnetic field $\boldsymbol{B}$ are the same as those in \citetalias{Bambic2024}, we must impose an additional boundary condition on $\boldsymbol{F}_{\rm con}$. We choose a simple outflow boundary condition, where the conductive heat flux is copied into the ghost zones with zero gradient. Note that this boundary condition is inconsistent with the condition on temperature, which is simply copied into the ghost zones from the last live zone such that $T(x,y,\pm z_{\rm ghost})=T(x, y, \pm z_{\rm max})$, where $\pm z_{\rm ghost}$ are the $z$-coordinates in the ghost zones. We find, though, that the choice of a simple zero-gradient boundary condition ensures the correct behaviour of the heat flux near the upper/ lower boundaries. 

The influence of magnetic fields is described in terms of the net vertical magnetic flux threading the disc, $\Phi_{\rm B} = \oiint \boldsymbol{B} \bcdot \hat{\bb{z}} \: \rmd x \rmd y$. For the zero-net-flux (ZNF) simulations, we initialize the field of \cite{Miller2000}, and for the net flux (NF) simulations, we initialize a constant vertical magnetic field, $B_z$. The magnetic-field strength is determined by the initial mid-plane plasma $\beta$ parameter at time $t=0$: $\beta_0 \equiv 2 P_0 (t=0)/|\boldsymbol{B}_0 (t = 0)|^2$. For the ZNF simulations, we study cases where $\beta_0 = 10^2$ (weak ZNF) and $10$ (moderate ZNF); for NF fields, we analyze cases with $\beta_0 = 10^4$ (weak NF) and $10^3$ (moderate NF).

\subsection{Numerical Details} \label{sec:numerics}

Because the shearing box loses mass via winds, we inject material to maintain an approximately constant steady-state disc mass. This mass injection ensures that the the density below the corona remains high enough to keep cooling and heating roughly balanced throughout the disc. Cooling and mass injection are handled in the same way as was done in Paper I. Our cooling function~\eqref{eq:cooling_function} is evaluated through the exact integration method presented in \cite{Townsend2009}. Mass is injected at a rate $\dot{\rho}_{\rm src}$ according to the initial density profile to compensate for the mass-outflow rate of the wind, $\dot{M}_{\rm wind}$, 
\begin{equation} \label{eq:Mdot_wind}
    \dot{M}_{\rm wind} (t) = \int \dot{\rho}_{\rm src} \: \rmd x \rmd y \rmd z = \oint \rho u_z \: \rmd x\rmd y.
\end{equation}
When we inject mass into a given cell, we maintain the velocity $\boldsymbol{u}$ and temperature $T$ already in that cell such that we are injecting mass, momentum, and energy into the domain and maintaining the constant mass of the box, $M_{\rm box}$. These source terms as well as the cooling loss term are implemented in operator-split fashion.

\begin{table*} 
\renewcommand{\arraystretch}{1.1}
\small\addtolength{\tabcolsep}{-2pt}
\scalebox{1}{%
\begin{tabular}{l c c c c c c}     
\hline  
Simulation & $z_T$   & $t_{\rm evap}$  & $t_{\rm dep}$  & $t_{\rm thm}$ & $\langle \dot{E}_{\rm turb} \rangle_t$ & $\langle \dot{E}_{\rm cool}^{\rm cor} \rangle_t$\\
           & $(H_z)$ & ($100$ orbits)  & ($100$ orbits) & (orbits)      & $(\rho_0 H_z^5 \Omega^3)$              & $(\dot{E}_{\rm turb})$ \\ 
\hline
$\rm{Conduction} \: \rm{NF} \: \: \: \: \mathcal{A} = 10 \: \: \: \beta_0 = 10^4$ & $1.6$ & $14 \pm 4$ & $2.5_{-0.9}^{+0.8}$ & $6.1_{-1.8}^{+1.7}$ & $4.2_{-0.9}^{+0.8}$ & $0.08 \pm 0.07$ \\
$\rm{Conduction} \: \rm{NF} \: \: \: \: \mathcal{A} = 10 \: \: \: \beta_0 = 10^4 \: \rm{HiRes}$ & $1.5$ & $10 \pm 2$ & $1.9 \pm 0.6$ & $5.1 \pm 0.5$ & $4.7_{-0.5}^{+0.6}$ & $0.04 \pm 0.02$ \\
$\rm{No} \: \rm{Conduction} \: \rm{NF} \: \: \: \: \mathcal{A} = 10 \: \: \: \beta_0 = 10^4$ & $1.7$ & $\mbox{---}$ & $2.1_{-0.8}^{+0.7}$ & $5.2 \pm 0.6$ & $4.7 \pm 0.7$ & $0.05 \pm 0.03$ \\
$\rm{Conduction} \: \rm{NF} \: \: \: \mathcal{A} = 10^3 \: \:  \beta_0 = 10^4$ & $2.8$ & $138 \pm 14$ & $7.3_{-2.6}^{+2.3}$ & $7.1 \pm 0.7$ & $3.3 \pm 0.4$ & $0.17 \pm 0.05$ \\
$\rm{No} \: \rm{Conduction} \: \rm{NF} \: \: \: \mathcal{A} = 10^3 \: \:  \beta_0 = 10^4$    & $3.5$ & $\mbox{---}$ & $6.5_{-1.9}^{+1.8}$ & $5.4_{-1.1}^{+1.2}$ & $3.9 \pm 0.7$ & $0.17_{-0.07}^{+0.06}$ \\
\hline 
$\rm{Conduction} \: \rm{NF} \: \: \: \: \mathcal{A} = 10 \: \: \: \beta_0 = 10^3$ & $2.5$ & $3 \pm 0.2$ & $0.4 \pm 0.1$ & $1.6 \pm 0.3$ & $22.0_{-4.3}^{+4.0}$ & $0.19_{-0.06}^{+0.05}$ \\
$\rm{Conduction} \: \rm{NF} \: \: \: \: \mathcal{A} = 10 \: \: \: \beta_0 = 10^3 \: \rm{HiRes}$ & $1.9$ & $2 \pm 0.2$ & $0.4 \pm 0.1$ & $1.0 \pm 0.2$ & $25.0_{-4.0}^{+5.0}$ & $0.20_{-0.05}^{+0.06}$ \\
$\rm{No} \: \rm{Conduction} \: \rm{NF} \: \: \: \: \mathcal{A} = 10 \: \: \: \beta_0 = 10^3$ & $2.4$ & $\mbox{---}$ & $0.4 \pm 0.1$ & $1.5 \pm 0.4$ & $22.0_{-3.8}^{+3.5}$ & $0.22_{-0.08}^{+0.07}$ \\
$\rm{Conduction} \: \rm{NF} \: \: \: \mathcal{A} = 10^3 \: \:  \beta_0 = 10^3$    & $4.1$ & $152 \pm 11$ & $0.8 \pm 0.3$ & $1.5 \pm 0.3$ & $20.1 \pm 2.7$ & $0.31_{-0.20}^{+0.21}$ \\
$\rm{No} \: \rm{Conduction} \: \rm{NF} \: \: \: \mathcal{A} = 10^3 \: \:  \beta_0 = 10^3$    & $4.2$ & $\mbox{---}$ & $0.7 \pm 0.2$ & $1.7 \pm 0.4$ & $20.7_{-3.3}^{+3.1}$ & $0.34 \pm 0.08$ \\
\hline 
$\rm{Conduction} \: \rm{ZNF} \: \: \: \: \mathcal{A} = 10 \: \: \: \beta_0 = 10^2$ & $1.8$ & $9 \pm 0.1$ & $212 \pm 100$ & $93_{-16}^{+18}$ & $0.3 \pm 0.04$ & $0.15_{-0.03}^{+0.02}$ \\
$\rm{No} \: \rm{Conduction} \: \rm{ZNF} \: \: \: \: \mathcal{A} = 10 \: \: \: \beta_0 = 10^2$ & $1.7$ & $\mbox{---}$ & $15_{-4}^{+5}$ & $35_{-6}^{+7}$ & $1.7 \pm 0.2$ & $0.08 \pm 0.01$ \\
$\rm{Conduction} \: \rm{ZNF} \: \: \: \mathcal{A} = 10^3 \: \:  \beta_0 = 10^2$    & $2.8$ & $253 \pm 2$ & $758_{-640}^{+690}$ & $39_{-14}^{+15}$ & $0.5 \pm 0.3$ & $0.22_{-0.10}^{+0.11}$ \\
$\rm{No} \: \rm{Conduction} \: \rm{ZNF} \: \: \: \mathcal{A} = 10^3 \: \:  \beta_0 = 10^2$    & $1.7$ & $\mbox{---}$ & $15_{-4}^{+5}$ & $25_{-2}^{+3}$ & $1.7 \pm 0.2$ & $0.08 \pm 0.01$ \\
\hline 
$\rm{Conduction} \: \rm{ZNF} \: \: \: \: \mathcal{A} = 10 \: \: \: \beta_0 = 10$ & $1.7$ & $8 \pm 2$ & $83_{-32}^{+31}$ & $20_{-3}^{+4}$ & $0.9 \pm 0.2$ & $0.09 \pm 0.02$ \\
$\rm{Conduction} \: \rm{ZNF} \: \: \: \: \mathcal{A} = 10 \: \: \: \beta_0 = 10 \: \rm{HiRes}$ & $1.7$ & $7 \pm 1$ & $63_{-25}^{+31}$ & $17 \pm 2$ & $1.0_{-0.1}^{+0.2}$ & $0.06 \pm 0.02$ \\
$\rm{No} \: \rm{Conduction} \: \rm{ZNF} \: \: \: \: \mathcal{A} = 10 \: \: \: \beta_0 = 10$ & $1.7$ & $\mbox{---}$ & $13 \pm  3$ & $12_{-2}^{+3}$ & $1.9 \pm 0.2$ & $0.07 \pm 0.01$ \\
$\rm{Conduction} \: \rm{ZNF} \: \: \: \mathcal{A} = 10^3 \: \:  \beta_0 = 10$    & $2.8$ & $257 \pm 32$ & $630_{-410}^{+320}$ & $59_{-9}^{+10}$ & $0.3 \pm 0.1$ & $0.28_{-0.06}^{+0.05}$ \\
$\rm{No} \: \rm{Conduction} \: \rm{ZNF} \: \: \: \mathcal{A} = 10^3 \: \:  \beta_0 = 10$    & $2.9$ & $\mbox{---}$ & $129_{-53}^{+48}$ & $21_{-5}^{+4}$ & $1.4 \pm 0.2$ & $0.11 \pm 0.02$ \\
\hline 
\end{tabular}}
\caption{Summary of key diagnostics for all 19~simulations. Quantities are quoted with $1 \sigma$ error bars, where the errors are computed from the 16th and 84th percentiles of the respective time series from 30 to 100 orbits in net flux (NF) simulations, from 30 to 50 orbits in `$\rm{No} \: \rm{Conduction}$' zero net flux (ZNF) simulations, and from 70 to 100 orbits in the `$\rm{Conduction}$' ZNF runs. Horizontally averaged temperature first rises above the mid-plane temperature $T_0$ at height $z_T$. The box-averaged turbulent injection power $\dot{E}_{\rm turb}$ is measured in code units, while the cooling within the corona ($\dot{E}_{\rm cool}^{\rm cor}$) is measured relative to $\dot{E}_{\rm turb}$.
} 
\label{table:simulations} 
\end{table*}

Our simulations are advanced in time using second-order-accurate van Leer time integration \citep[\texttt{vl2};][]{vanLeer1979} and spatial integration is performed using the Harten--Lax--van Leer Discontinuity (\texttt{HLLD}) Riemann solver with second-order-accurate piecewise-linear-method (\texttt{PLM}) reconstruction. As in Paper I, we do not use orbital advection, and we use the smoothed vertical gravitational potential from \cite{Davis2010} (see \citetalias{Bambic2024} Equation 23).

Simulations with conduction are more expensive than those without conduction because a large value of $v_{\rm max}$ must be chosen to ensure that the equilibrium heat flux is realized, i.e., to satisfy equation~\ref{eq:vmax_condition}. This large signal speed must be resolved by a short time step. To cut down computational costs, we use $\rho_{\rm floor}/ \rho_0 = 10^{-4}$ for the NF simulations and $\rho_{\rm floor}/ \rho_0 = 10^{-6}$ for the ZNF simulations, with the pressure floor set as $P_{\rm floor} = \rho_{\rm floor} T_0$. Further, while we use the same domain size as \citetalias{Bambic2024}, i.e., $(L_x, L_y, L_z) = (4 H_z, 8 H_z, 12 H_z)$, which allows for a comparison of these conduction runs with our earlier results, we run the more expensive conduction simulations at half \citetalias{Bambic2024}'s resolution: 16 grid cells per $H_z$. We perform higher resolution, 32 cells/$H_z$, simulations of three of our models: the moderate NF $\mathcal{A} = 10$, weak NF $\mathcal{A} = 10$, and moderate ZNF $\mathcal{A} = 10$ models, to verify that our conclusions do not depend on the chosen resolution. All simulations are evolved for 100 orbits until time $\Omega t \approx 628$.

\subsection{Choice of $v_{\rm max}$} \label{sec:vmax}

The two-moment method used to evolve the conductive heat flux requires a choice of the maximum signal speed $v_{\rm max}$, which sets the size of the time-derivative term in the evaluation of the heat flux. This term is negligible so long as Equation~\ref{eq:vmax_condition} is satisfied. To determine a sensible choice for $v_{\rm max}$, we leveraged intuition gained from the simulations in \citetalias{Bambic2024} to realize that, in those simulations, the Courant condition is set by the large \alf speed in the corona: $v_{\rm A} \equiv |\boldsymbol{B}|/\sqrt{\rho}$. Using the simulation data from \citetalias{Bambic2024}, we found that the horizontally averaged \alf velocity peaks at around $5 c_{s,0}$ in the corona, i.e., ${\approx}3.5 \: H_z \Omega$. We choose a maximum velocity that greatly exceeds this averaged \alf velocity, $v_{\rm max} = 30 \: H_z \Omega$, to ensure accurate computation of the heat flux. We further explore the effect of $v_{\rm max}$ on our conclusions in Appendix~\ref{sec:vmax_appendix}. 

\subsection{Simulations and Diagnostics} \label{sec:diagnostics}

We run a series of 19 simulations as detailed in Table~\ref{table:simulations}. These simulations scan over two values of the Coulomb coupling parameter $\mathcal{A} \in \{ 10, 10^3 \}$ and four different field configurations: weak NF ($\beta_0~=~10^4$), moderate NF ($\beta_0=10^3$), weak ZNF ($\beta_0=10^2$), and moderate ZNF ($\beta_0=10$). For each combination of $\mathcal{A}$ and field configuration, we provide an analysis of runs with and without field-aligned, free-streaming thermal conduction. 

For any simulations with NF, we run full simulations from 0 to 100 orbits with thermal conduction either always on or always off. However, in simulations initialized with ZNF, we found that early on at times $t < 5$ orbits, conduction acted to collapse the upper atmosphere onto the disc before a corona could properly form. This collapse evacuated the atmosphere above $|z| = 2 H_z$, resulting in the continual addition of floored material at large $|z|$. Subsequently, the timestep dropped dramatically. To avoid this initial collapse, we run all ZNF simulations from 0 to 50 orbits with conduction effectively turned off at the level of $\kappa = \kappa_{\rm min}$ (denoted as `$\rm{No} \: \rm{Conduction}$' in Table~\ref{table:simulations}). Then, at $t = 50$ orbits, conduction is turned on. As we show in \S\ref{sec:condensing_inflows}, a condensing inflow onto the disc forms, and a steady state is reached by 70 orbits. Thus, a single ZNF simulation provides a `$\rm{No} \: \rm{Conduction}$' run between 0 and 50 orbits and a `$\rm{Conduction}$' run from 50 to 100 orbits. All of our simulations are run at a resolution of 16 grid cells/$H_z$ with the exception of the 3 runs denoted as `$\rm{HiRes}$' in Table~\ref{table:simulations}, which are run at a resolution of 32 cells/$H_z$.

\begin{figure*}
\hbox{
\includegraphics[width=1.0\textwidth]{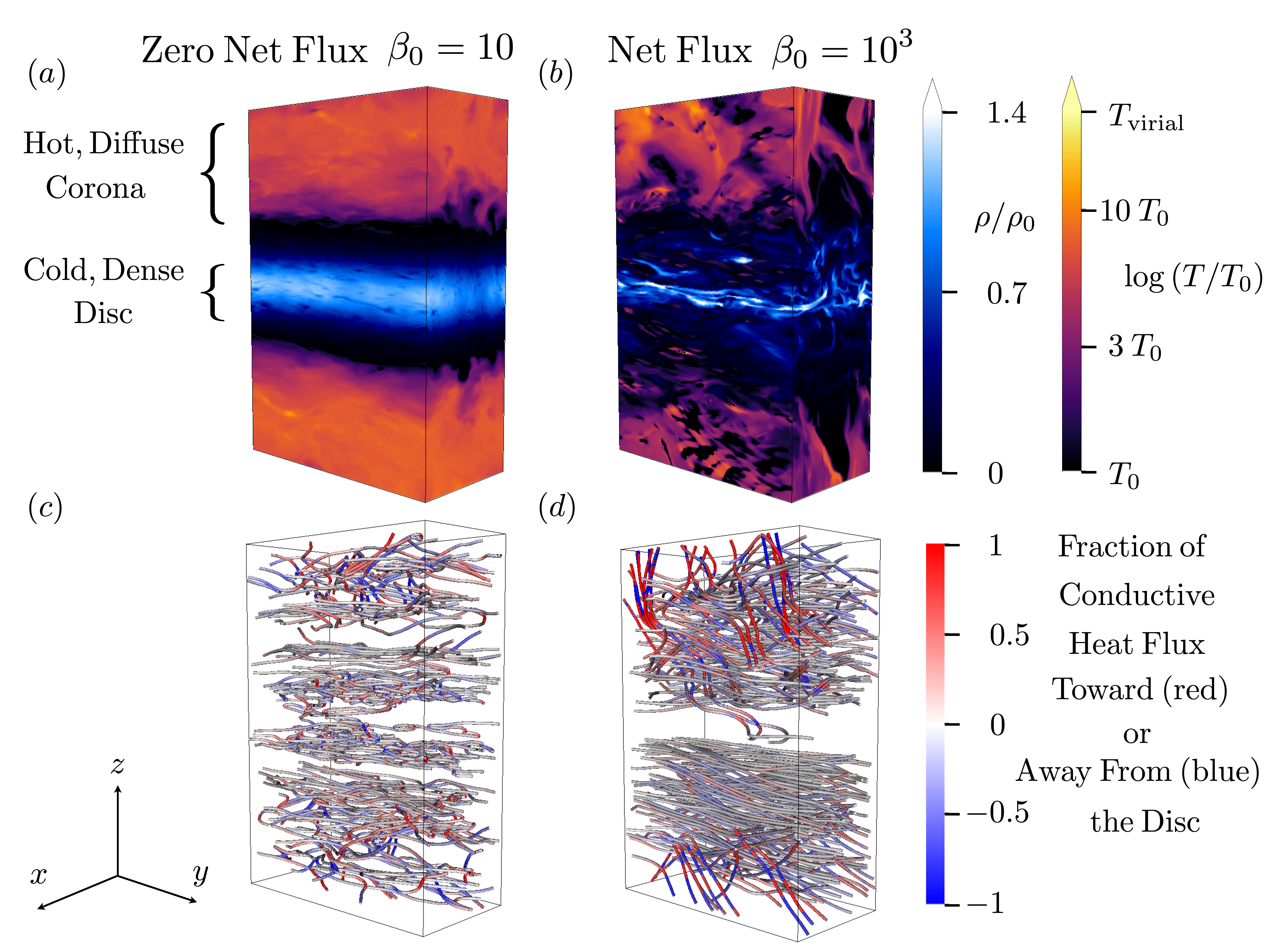}
}
\caption{Top row: Volume renderings of temperature ($T$) and density ($\rho$) for the zero-net-flux (ZNF), $\mathcal{A} = 10$, $\beta_0 = 10$ (panel a) and net-flux (NF), $\mathcal{A} = 10$, $\beta_0 = 10^3$ (panel b) simulations at times $t = 80$ orbits and $t = 46.2$ orbits, respectively. Bottom row: Volume renderings of geometric suppression factor $f_s \equiv F_{{\rm con}, z}/|\boldsymbol{F}_{\rm con}|$ for the ZNF, $\mathcal{A} = 10$, $\beta_0 = 10$ (panel c) and NF, $\mathcal{A} = 10$, $\beta_0 = 10^3$ (panel d) simulations. In panels (a) and (b), temperature is only shown where $T > 1.2 T_0$, with $T_0$ as the mid-plane temperature; density is shown where temperature is below this threshold. While temperature inversions are present in both simulations, the vertical structure of the discs are unaffected by ion thermal conduction, even at the free-streaming limit.
}
\label{fig:volume_renderings}
\end{figure*}

\begin{figure*}
\hbox{
\includegraphics[width=1.0\textwidth]{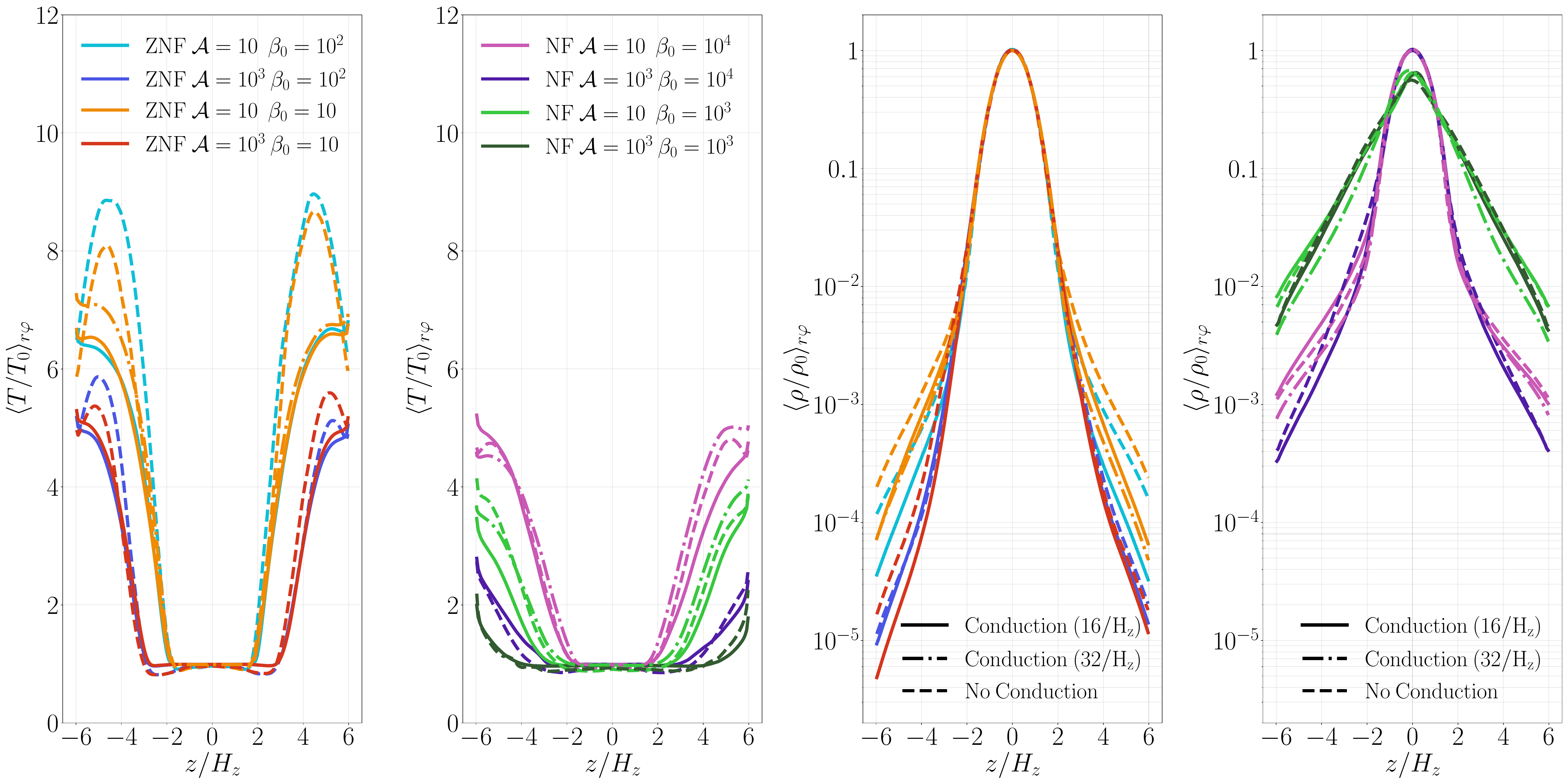}
}
\caption{Left: Horizontally averaged temperature ($T$) profiles from all simulations. Profiles for net-flux simulations with and without conduction are averaged over 30--100 orbits, while zero-net-flux simulations are averaged over two different time frames: 30--50 orbits for `No Conduction' runs and 70--100 orbits for `Conduction' runs. Right: Horizontally averaged density ($\rho$) profiles for all simulations over the same time frames. Profiles from both the standard-resolution (16 grid zones per $H_z$) and high-resolution (32 grid zones per $H_z$) simulations are shown. All simulations display temperature inversions, with a hot corona surrounding a cold disc, though runs without conduction (dashed lines) generally show higher temperatures and larger density enhancements relative to those with conduction (solid and dash-dot lines). Decreased densities and temperatures in the ZNF simulations with conduction relative to runs without conduction are the result of the coronae condensing onto the discs once conduction is activated at 50 orbits. 
}
\label{fig:temperature_density}
\end{figure*}

Throughout this work, we make use of the time average, $\langle \,\dots \rangle_t$, which represents the average from time $t_i$ to $t_f$. We choose $(t_i, t_f) = (30, 100)$ orbits in the NF simulations, $(t_i, t_f) = (30, 50)$ orbits in the ZNF `$\rm{No} \: \rm{Conduction}$' runs, and $(t_i, t_f) = (70, 100)$ orbits in the ZNF `$\rm{Conduction}$' simulations. Further, we introduce the horizontal and vertical averages as $\langle\,\dots \rangle_{r \varphi}$ and $\langle \,\dots \rangle_z$, which are averages over the $x-y$ plane and over the extent of the box in $z$, respectively. The notation $\langle \boldsymbol{F}_{\rm con} \rangle_{r \varphi z t}$ is shorthand for $\langle \langle \langle \boldsymbol{F}_{\rm con} \rangle_{r \varphi} \rangle_{z} \rangle_t$.

To assess if conductive evaporation can mediate the destruction of the accretion disc, we define an evaporation time $t_{\rm evap}$ as 
\begin{equation}
    t_{\rm evap} = \frac{M_{\rm box}}{\dot{M}_{\rm evap}} = \frac{M_{\rm box}}{- \frac{\gamma - 1}{\gamma} \frac{1}{T(z_F)} \oiint \boldsymbol{F}_{\rm con} (z_F) \bcdot \hat{z} \: dx dy},
\end{equation}
where the surface integral evaluates the peak conductive flux into the disc measured at the point where the flux into the disc is maximal, a height that we denote $z_F$. Note that $|z_F|$ is not necessarily the same above the below the mid-plane due to symmetry breaking in the shearing box. This definition of the evaporation time $t_{\rm evap}$ measures the evaporative outflow rate $\dot{M}_{\rm evap}$ for the maximum time-averaged heat flux into the disc---a `best case scenario' for the timescale over which the disc can evaporate \textit{in the absence of cooling}. Similarly, the wind depletion time is given by
\begin{equation} \label{eq:tdep}
    t_{\rm dep} \equiv \frac{M_{\rm box}}{\dot{M}_{\rm wind}}.
\end{equation} 

Energy is injected into the domain at the rate \citep{Hawley1995}
\begin{equation} \label{eq:turbulent_injection}
    \dot{E}_{\rm turb} = \int Q^{+}_{\rm turb} \: \rmd^3 \boldsymbol{r} = \frac{1}{2} q \Omega L_x \oiint \mathcal{T}_{r \varphi} (x = \pm L_x/2) \: \rmd y \rmd z,
\end{equation}
where $q \equiv -\rmd \ln{\Omega}/\rmd \ln{r} = 3/2$ is the shear parameter for a Keplerian flow, $L_x$ is the radial extent of the domain, and the surface integral is taken over the shearing boundary. Energy injection is modulated by the $r$-$\varphi$ component of the turbulent stress tensor,
\begin{equation} \label{eq:Trphi}
    \boldsymbol{\mathcal{T}}_{r \varphi} \equiv \rho \boldsymbol{v} \boldsymbol{v} - \boldsymbol{B} \boldsymbol{B},
\end{equation}
where the turbulent velocity $\boldsymbol{v}= \boldsymbol{u} + \frac{3}{2} \Omega x \hat{\bb{y}}$ does not include the background Keplerian shear.

Thermal equilibrium, i.e., a balance of heating and cooling, is established in the disc on the thermal timescale,
\begin{equation} \label{eq:tthermal}
    t_{\rm thm} \equiv \frac{P_0}{\langle \mathcal{T}_{r \varphi} \rangle_{r \varphi t} (z = 0)} \frac{1}{\Omega} = \frac{1}{\alpha_{\rm mid} \Omega},
\end{equation}
where we have introduced the \cite{Shakura1973} `$\alpha$' parameter, which we define in terms of the mid-plane pressure $P_0$ such that $\alpha \equiv \langle \mathcal{T}_{r \varphi}/P_0 \rangle_{r \varphi t}$. We define $\alpha_{\rm mid} \equiv \alpha (|z| < 2 H_z)$. Note that this definition of $\alpha_{\rm mid}$ is different than that used in \citetalias{Bambic2024}. For the lower resolution 16/$H_z$ simulations explored in this work, we find that there is increased numerical dissipation at the mid-plane, resulting in a substantial drop in $\alpha$ within $|z| < H_z/2$, primarily for the ZNF simulations. To compensate for this resolution-induced drop in $\alpha$, we choose to measure the thermal time based on the range $|z| < 2 H_z$. Note that at higher 32/$H_z$ resolution, even in runs with conduction, the thermal times are similar to the comparable simulations in Paper~I for the same combination of $\mathcal{A}$ and field configuration. 

Finally, the cooling rate in the corona is measured by evaluating the following integral, 
\begin{equation} \label{eq:cooling_rate}
    \dot{E}_{\rm cool}^{\rm cor} = \int Q_{\rm cool}^{\rm cor} \: \rmd^3 \boldsymbol{r} = \int \Lambda (|z| > 2 H_z) \: \rmd^3 \boldsymbol{r},
\end{equation}
where the volume integral is taken over the entire corona. Following \citetalias{Bambic2024}, we define the corona as the region $|z| > 2 H_z$ above the disc mid-plane. In a simulation with radiation transport, the corona is self-consistently defined by the surface above which the optical depth drops to $\tau_{\rm es} \leq 1$. Since our simulations do not include radiation transport, a more appropriate definition for the corona may be the region above $z_T$, where $z_T$ is the height at which the temperature begins to rise rapidly above $T_0$. Table~\ref{table:simulations} lists $z_T$ for all simulations. Since $z_T > 2.5 H_z$ in strongly cooled ($\mathcal{A} = 10^3$) simulations, the cooling rates in the corona ($|z| > 2 H_z$) may be over-estimated. Ultimately, radiation transport simulations are necessary to evaluate the amount of cooling in the optically thin corona vs. the optically thick disc. However, evaporation or survival of a cold $T \leq T_0$ accretion disc is independent of the definition of the corona. 

\begin{figure*}
\hbox{
\includegraphics[width=1.0\textwidth]{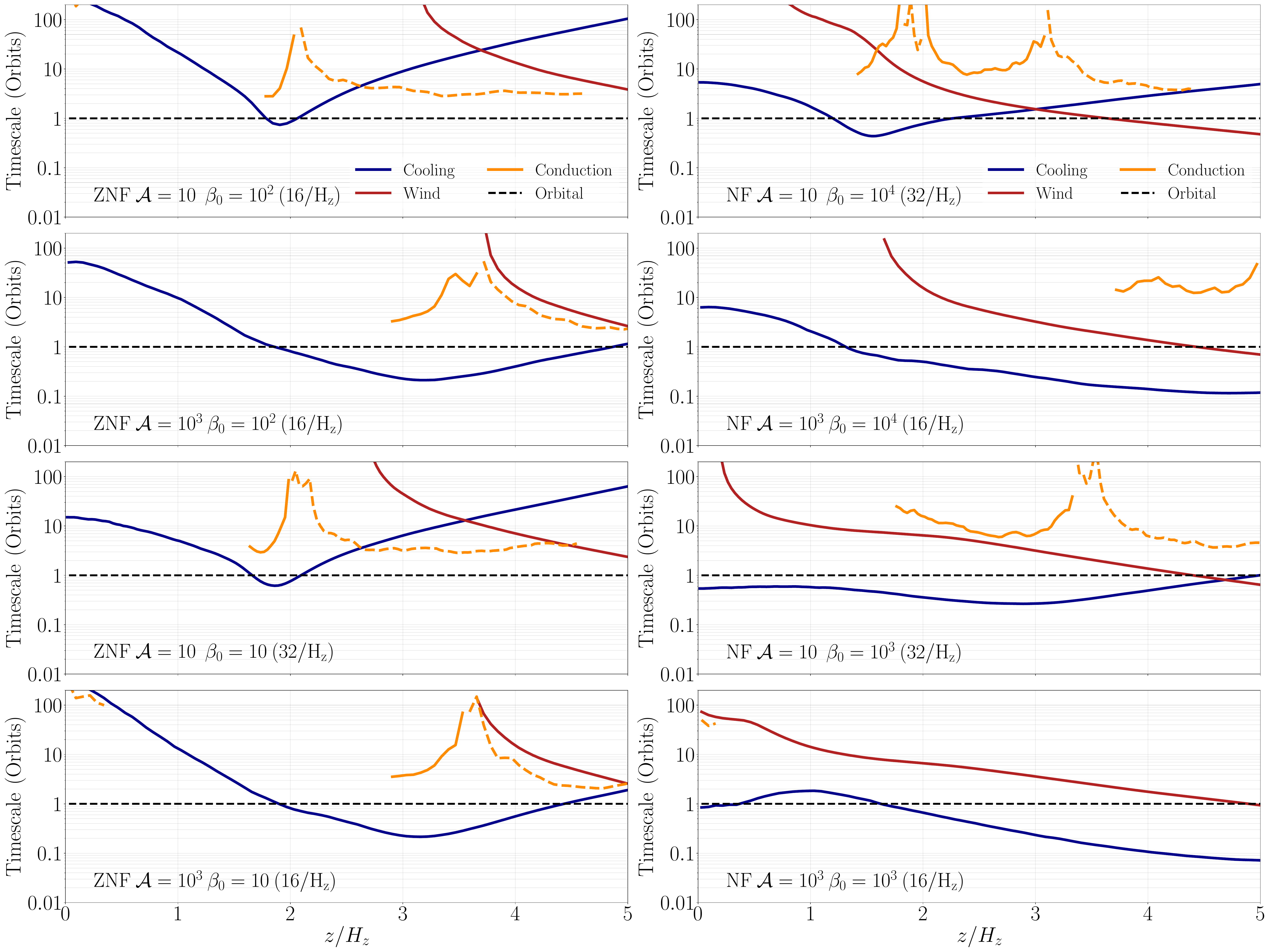}
}
\caption{Timescale profiles for all simulations with thermal conduction. We compare the conductive heating timescale (Equation~\ref{eq:tcond}; orange, solid) and conductive cooling timescale (Equation~\ref{eq:tcond}; orange, dashed) to the wind outflow time $t_{\rm wind}$ (red; Equation~\ref{eq:t_wind}), the measured cooling time $t_{\rm cool}$ (blue; Equation~\ref{eq:tcool}), and the orbital timescale $t_{\rm orb}$ (black dashed). For zero-net-flux cases (left column), conductive cooling is important relative to two-temperature cooling above $z = 3 H_z$, leading to modest condensation of the corona onto the disc. Conductive heating, on the other hand, is always subdominant to two-temperature cooling. In the net flux simulations (right column), the conductive heating and cooling timescales are always longer than the wind-outflow timescales.
}
\label{fig:timescales}
\end{figure*}

\section{Results} \label{sec:results}

\subsection{The absence of disc evaporation} \label{sec:no_evaporation}

The accretion disc does not evaporate in any of our simulations, independent of the presence of net vertical magnetic flux and the strength of Coulomb coupling in the corona, as parameterized by $\mathcal{A}$. In what follows, we show quantitatively that the coronal structure and outflow rates do not change in the presence of conduction, consistent with no significant evaporation. 

Figure~\ref{fig:volume_renderings} illustrates the temperature, density, and magnetic-field structures in our simulations. All of our simulations form temperature inversions, with a hot, diffuse corona `sandwiching' a colder, thin disc (see panels~a and~b of Fig.~\ref{fig:volume_renderings}). These inversions establish a temperature gradient that channels the conductive heat flux on average toward the disc mid-plane at heights above $z_T$. Conductive heat fluxes are directed along magnetic-field lines (panels~c and~d in Fig.~\ref{fig:volume_renderings}), and the vertical heat flux into the disc is suppressed by the toroidal fields formed via the strong Keplerian shear. Simulations with NF, particularly when the field is of moderate strength ($\beta_0 = 10^3$), exhibit well-ordered magnetic fields, including many nearly vertical field lines threading the corona yet anchored into the disc. Heat fluxes along these field lines are largely uninhibited. 

Yet, in spite of the heat flux from the corona, discs with mid-plane temperatures $T \leq T_0$ persist effectively unchanged in all simulations. This feature of our simulations is demonstrated by the horizontally and time averaged temperature and density profiles in Figure~\ref{fig:temperature_density}.  

Examining the depletion times in Table~\ref{table:simulations}, we find that conduction has no statistically significant effect on the depletion times, i.e., the mass-outflow rates are not enhanced in any meaningful way by an evaporative outflow. In NF runs, $t_{\rm dep}$ values from simulations with and without conduction agree within the $1 \sigma$ error bars when comparing simulations with the same initial field configuration and the same $\mathcal{A}$. Even though simulations with NF launch strong magnetocentrifugally driven outflows (see \citetalias{Bambic2024}), these outflows are in no way fed by a conduction-induced evaporative flow. 

A key goal of this paper is to understand which terms in Equation~\ref{eq:evaporation_criterion} are responsible for the lack of evaporation in our models, and if we should expect these results to carry over to a realistic, global system. In the subsections that follow, we explore (1) the steepness of the temperature gradient, as measured by $|L_T|$ (\S\ref{sec:temperature_inversions}), (2) the rate of Coulomb cooling relative to conductive heating and advection by an outflow in the corona (\S\ref{sec:timescales}), and (3) the role of the magnetic field in geometrically suppressing the heat flux into the disc (\S\ref{sec:geometric_suppression}).

\subsection{Temperature inversions and conductive cooling} \label{sec:temperature_inversions}

Figure~\ref{fig:temperature_density} displays the horizontally and temporally averaged temperature and density profiles for all simulations with and without conduction. All ZNF simulations form temperature inversions, with a hot corona forming above $|z| = 3 H_z$. Conduction has no effect on the location of $z_T$, the height at which the horizontally averaged temperature profile first rises above $T_0$. This feature indicates a lack of substantial conductive heating in the disc's surface layers. We find minimum values of $|L_T|/H_{z, {\rm c}}$ in the range of $0.1 < |L_T|/H_{z, {\rm c}} < 0.5$ for the ZNF simulations and $0.9 < |L_T|/H_{z, {\rm c}} < 1.5$ for the NF simulations. The effect of conduction is to increase $|L_T|/H_{z, {\rm c}}$, but only by a few tens of per cent relative to runs without conduction. 

Rather than heating the disc and moving $z_T$ inward, conduction in ZNF simulations acts primarily to cool the coronae. The result is a 25 per cent drop in temperature in the weakly cooled, $\mathcal{A} = 10$ ZNF runs. Drops in temperature precipitate decreases in the coronal density as material condenses onto the disc; however, these changes in density structure likely have a negligible observable effect. At $|z| = 2 H_z$, all ZNF simulations have approximately the same horizontally averaged densities. Above $|z| = 4 H_z$, the average coronal density can drop by as much as a factor of 2 once conduction is turned on. This drop is more substantial for the $\mathcal{A} = 10$ runs, indicating that the density drop is simply due to a decrease in thermal pressure support caused by conductive cooling.

In NF simulations, temperature and density profiles with and without conduction are nearly indistinguishable from one another. The only notable feature, present in both the ZNF and NF simulations, is that in runs without conduction, there is a characteristic peak feature in the temperature profiles above $|z| = 4 H_z$ and an outward decrease in temperature above this peak. This feature is a boundary effect; the box-height scan in Appendix~B of Paper I demonstrates that the peak moves outward with increasing box height. With conduction, the temperature continues to rise monotonically into the boundary and no peak is evident. At these large heights above the disc, the plasma is quite diffuse, and conduction dominates over cooling enough to modify the thermal structure of the corona.

\subsection{Thermodynamic timescales} \label{sec:timescales}

Conduction redistributes heat amongst different heights in the accretion disc and corona. We define the conductive heating/cooling timescale as
\begin{equation} \label{eq:tcond}
    t_{\rm cond} \equiv \frac{1}{\gamma - 1} \frac{\langle P \rangle_{r \varphi t}}{\left|\frac{\partial}{\partial z} \langle F_{{\rm con}, z} \rangle_{r \varphi t}\right|},
\end{equation}
where $F_{{\rm con}, z} = \boldsymbol{F}_{\rm con} \bcdot \hat{z}$ is the conductive heat flux in the $z$-direction. When $\partial/\partial z \: \langle F_{{\rm con},z} \rangle_{r \varphi} < 0$, the above expression refers to the conductive heating timescale, and when the gradient in the heat flux is positive, $t_{\rm cond}$ is the conductive cooling timescale. The conductive heating/ cooling timescale can then be compared to the cooling time,
\begin{equation} \label{eq:tcool}
    t_{\rm cool} (z) \equiv \frac{3}{2} \frac{\langle P \rangle_{r \varphi t}}{\langle Q_{\rm cool}^{-} \rangle_{r \varphi t}},
\end{equation}
and the wind-outflow timescale, i.e. the timescale over which the outflow can transport plasma over a height $2 H_z$, 
\begin{equation} \label{eq:t_wind}
    t_{\rm wind} (z) \equiv \frac{2 H_z}{|\langle v_z \rangle_{r \varphi t}|}.
\end{equation}

Figure~\ref{fig:timescales} displays a comparison of these different timescales. In the ZNF simulations, conductive heating (solid orange line in Figure~\ref{fig:timescales}) acts only over a very narrow range of $z$, less than half a mid-plane scale height above $z_T$. Within this range, conductive heating is slow compared to cooling (blue curves), with the conductive heating timescale exceeding the cooling timescale by an order of magnitude in the strongly cooled, $\mathcal{A} = 10^3$ cases, and by a factor of a few in the weakly cooled, $\mathcal{A} = 10$ runs. 

Conductive heating occurs over a much broader range of heights in the NF simulations than in the ZNF simulations. The range of conductive heating extends from $z_T$ to ${\approx}(z_T + H_z)$ for all but the NF, $\mathcal{A} = 10^3$, $\beta_0 = 10^3$ run. As in the ZNF simulations, cooling outpaces conductive heating at all $z$. Unlike the ZNF simulations, though, the wind-outflow timescale is shorter than the conductive heating timescale in all NF runs, with the exception of a very narrow range near $z= z_T = 1.5 H_z$ in the NF, $\mathcal{A} = 10$, $\beta_0 = 10^4$ simulation. Hot plasma in the corona is transported upwards more rapidly than conduction can channel thermal energy down towards the disc. 

We can extrapolate these results to a global system by comparing the best-case scenario evaporation timescale, $t_{\rm evap}$, to the inflow (viscous) timescale of the disc, $t_{\rm visc} \equiv (R_0/H_z)^2 t_{\rm thm}$. Thinner discs slow the inward flow of plasma. In the \textit{absence} of two-temperature cooling, we can compute a critical disc thickness for conduction to evaporate the disc within the inflow time. The critical disc thicknesses to allow $t_{\rm evap} < t_{\rm visc}$ are $H_z/R_0 \approx (1-7) \times 10^{-2}$ for the NF simulations, and $H_z/R_0 \approx 0.05 - 0.3$ for the ZNF simulations. Particularly for the NF simulations, these discs are quite thin; however, such a thin disc may be realizable. This simple argument demonstrates that the lack of evaporation is not due to a negligible heat flux into the disc. Instead, Coulomb cooling radiates away the heat channeled toward the disc, preventing evaporation. 

Notably, while the conductive heating/cooling timescales are long compared to the two-temperature cooling time in all NF simulations, for weakly cooled ($\mathcal{A} = 10$) ZNF simulations, the conductive cooling timescale is actually \textit{shorter} than the cooling timescale imposed by Coulomb collisions above a height $z \geq 3 H_z$. Efficient conductive \textit{cooling} in the diffuse coronae of the ZNF simulations thus results in a collapse of the corona onto the disc in the form of a condensing inflow. We study the behaviour of this inflow further in \S\ref{sec:condensing_inflows}. 

\begin{figure}
\hbox{
\includegraphics[width=0.48\textwidth]{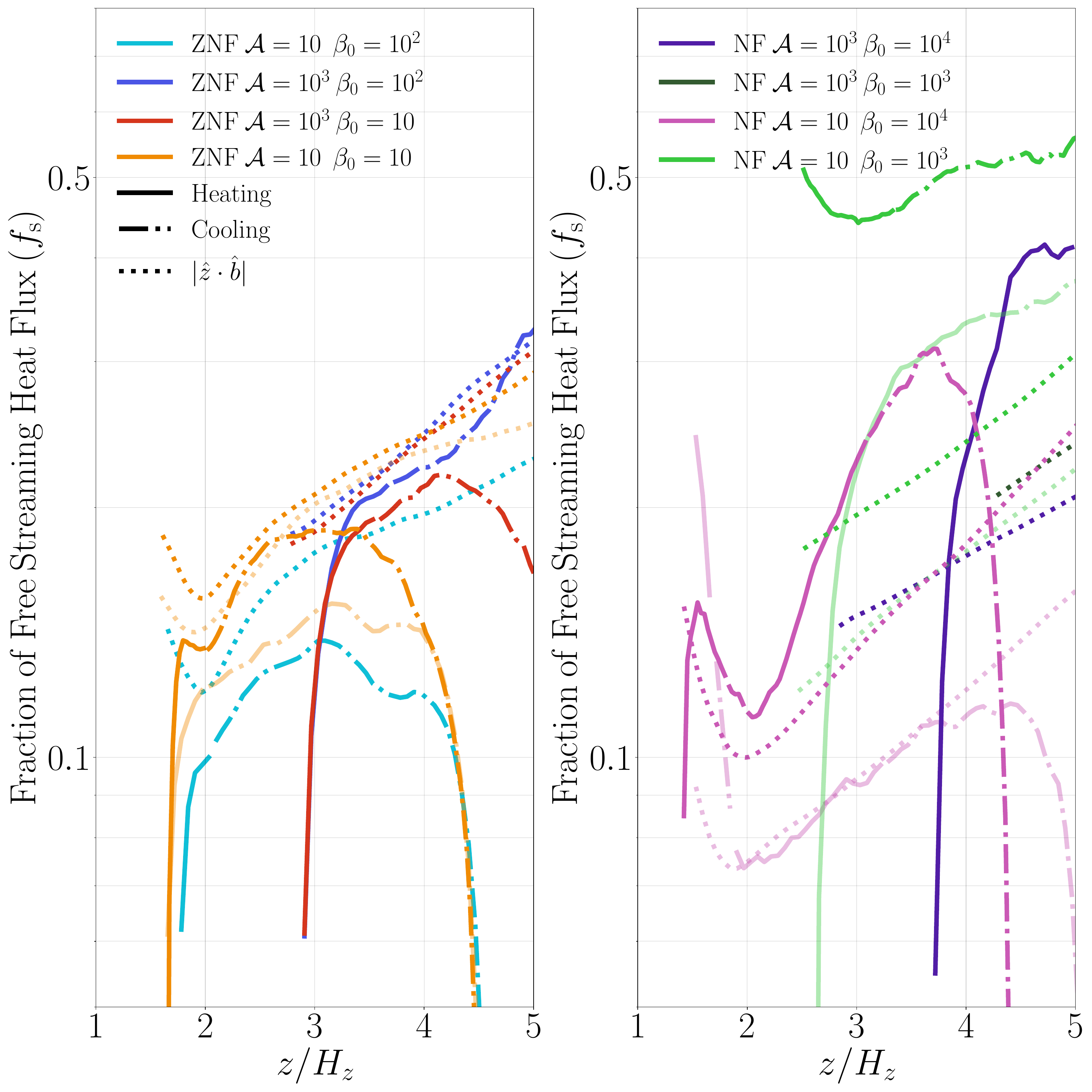}
}
\caption{Profiles of the conductive suppression factor $\langle f_{\rm s} \rangle_{r \varphi t}$ for all simulations, averaged from 30--100 orbits for the NF simulations, and averaged from 70--100 orbits in the ZNF cases. The suppression factor in regions where conduction is heating the plasma (solid) and cooling the plasma (dash-dotted) are shown separately, and a comparison to a na\"ive estimate for the suppression factor (viz., the horizontal average of $|\hat{z} \bcdot \hat{b}|$), is shown by the dotted lines. Where available, we use the high-resolution run for the profiles, but show the same result for the lower-resolution analogs using more transparent curves. We only plot the suppression factor in regions where $z \geq z_T$, as there is no consistent temperature gradient below this height.
}
\label{fig:fs_profiles}
\end{figure}

\subsection{Geometric suppression of the heat flux} \label{sec:geometric_suppression}

The vertical conductive heat fluxes are suppressed by the geometry of the predominantly toroidal magnetic field. To quantify this geometric suppression, we introduce the suppression factor $f_s$, and we define the horizontal average of the suppression factor as 
\begin{equation} \label{eq:f_suppression}
    \langle f_s \rangle_{r \varphi t} \equiv -\frac{\langle \boldsymbol{F}_{{\rm con}} \bcdot \hat{z} \rangle_{r \varphi t}}{\langle {\rm sgn} \left( \hat{z} \bcdot \grad T \right) |\boldsymbol{F}_{\rm con}| \rangle_{r \varphi t}}.
\end{equation}
Profiles of $\langle f_s \rangle_{r \varphi t}$ are shown in Figure~\ref{fig:fs_profiles}. A few features are noteworthy. First, the geometric suppression factors are generally lower in the ZNF simulations than those in the NF simulations. This trend is a result of the fact that fields with larger initial NF are more rigid, allowing them to remain more inclined with respect to the disc mid-plane and thus maintain a stronger coupling between the upper, hotter regions of the corona and cooler regions just above $z_T$. Generally, the suppression factor is in the range $0.1 < \langle f_s \rangle_{r \varphi t} < 0.2$ for the ZNF simulations, while a much larger range is spanned by the NF simulations:  $0.07 < \langle f_s \rangle_{r \varphi t} < 0.6$. Stronger cooling (larger $\mathcal{A}$) results in less suppression of the heat flux. This is because stronger cooling inhibits winds and evaporation of material into the corona. Thus, less mass is loaded on field lines, the field can maintain larger inclination angles, and heat fluxes can more easily pass among layers of the stratified flow. 

Interestingly, the magnitude of the suppression factor is resolution-dependent, and simulations performed at higher resolution exhibit \textit{less} suppression of the heat flux. At high resolution, we find suppression factors similar to those used by \citetalias{Cho2022}. Lower resolution runs, particularly the NF, $\mathcal{A} = 10$, $\beta_0 = 10^4$ simulation at 16 cells/$H_z$ resolution, display suppression factors well below these values.

Na\"ively using the field geometry as a proxy for the suppression factor by horizontally averaging the quantity $|\hat{z} \bcdot \hat{b}|$, i.e. the pitch angle cosine, shown as dotted lines in Figure~\ref{fig:fs_profiles}, also underestimates the heat flux by an order unity factor of ${\approx} 2-5$. The pitch angle cosine accounts for the geometry of the field, but not the direction of the heat flux, which is set by the temperature gradient. Because the flow is highly turbulent, the heat flux is not always directed toward the disc. This feature of the solutions is visible in panel~(d) of Fig.~\ref{fig:volume_renderings}. There, we see that the same field line can have neighbouring red regions (where the heat flux is toward the disc) and blue regions (where the heat flux is away from the disc), with a sharp transition between them, caused by a change in the direction of the temperature gradient along the field line. Failing to account for the direction of the gradient results in an underestimate of the suppression factor, and subsequently, an underestimate of the field-aligned heat flux into the disc. 

\subsection{Condensing inflows with zero net flux} \label{sec:condensing_inflows}

\begin{figure}
\hbox{
\includegraphics[width=0.48\textwidth]{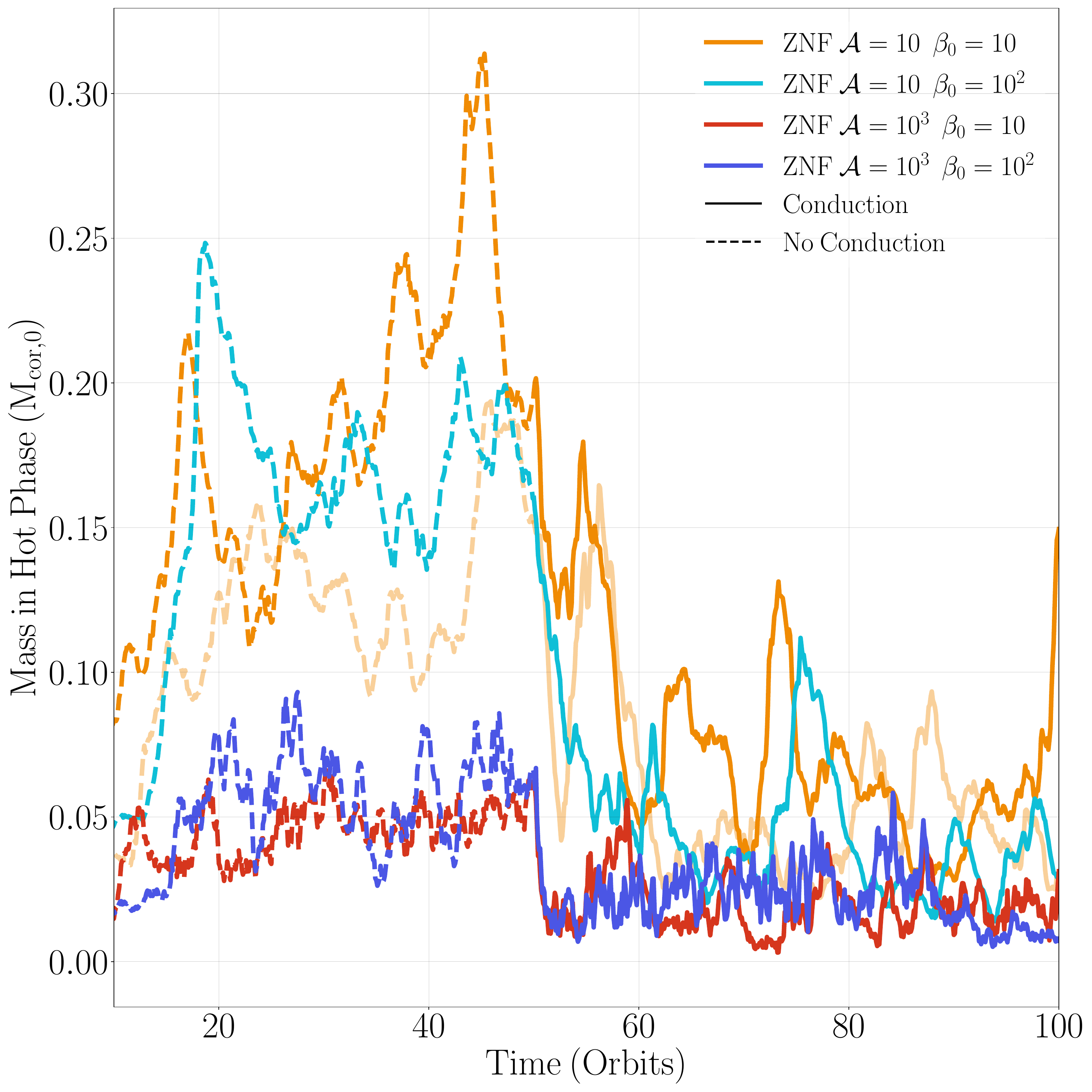}
}
\caption{Evolution of hot phase ($T \geq 5 T_0$) in the corona as a function of time for all ZNF simulations. The period when thermal conduction is turned off at the minimum value of $\kappa = \kappa_{\rm min}$ is indicated by the dashed portions of the curves while the period during which conduction is active at the free-streaming value is shown by the solid portion of the curves. We define the initial coronal mass ${\rm M}_{{\rm cor},0}$ as the total mass in the region $|z| \geq z_T$ from $t$~=~30--50 orbits, when conduction is not active. For the $\mathcal{A} = 10$ $\beta_0 = 10$ simulation, both the high resolution run (darker) and the fiducial resolution run (lighter/ more transparent) are shown. Conductive cooling produces a condensing flow from the corona onto the disc, reducing the mass of plasma in the hot phase by as much as 80 per cent when cooling is weak ($\mathcal{A} = 10$) and 50 per cent for much stronger two-temperature cooling ($\mathcal{A} = 10^3$), independent of the initial strength of the ZNF field. 
}
\label{fig:condensing_inflow}
\end{figure}

\begin{figure}
\hbox{
\includegraphics[width=0.5\textwidth]{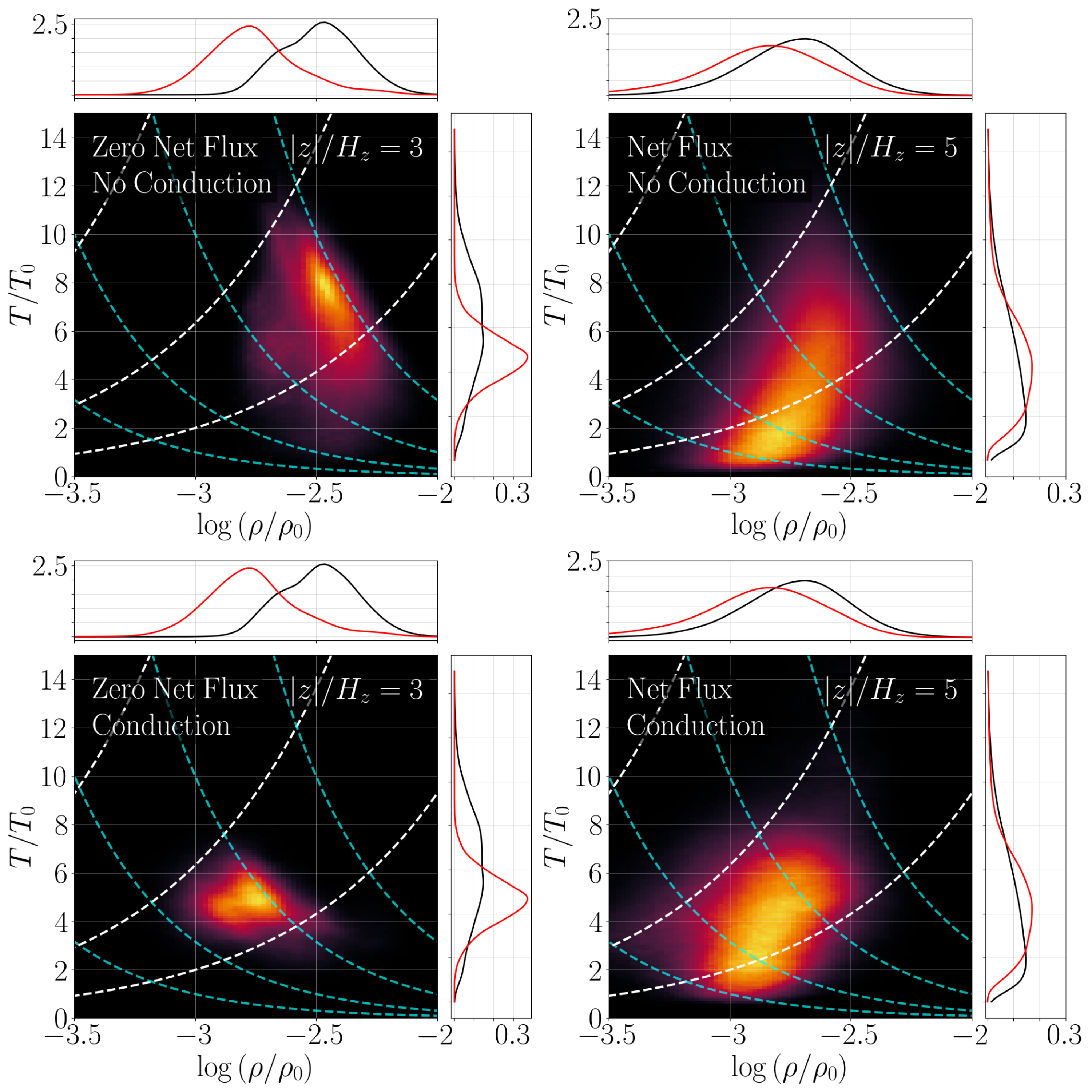}
}
\caption{Density-temperature distributions for all plasma in high resolution $\beta_0 = 10$ $\mathcal{A} = 10$ ZNF simulations at times 30--50 orbits (`No Conduction'; top left) and 70--100 orbits (`Conduction'; bottom left) at height $|z| = 3 H_z$ (left column). We show similar density-temperature distributions for all of the plasma in high resolution NF $\beta_0 = 10^4$ $\mathcal{A} = 10$ simulations at times 30--100 orbits for the simulation without conduction (`No Conduction'; top right) and the run with conduction (`Conduction'; bottom right) at height $|z| = 5 H_z$ (right column). White curves over the distributions represent lines of constant entropy, while light blue curves represent lines of constant pressure. We show one-dimensional probability density functions (PDFs) in $\log{(\rho/\rho_0)}$ and $T/T_0$ for all simulations, normalized such that the integrals under the curves equal 1, with black lines corresponding to runs without conduction, and red lines corresponding to simulations with conduction. Conduction reduces the spread in ion temperature within these multiphase coronae; however, the spread in density is largely unaffected, implying that conduction should not affect the optical depth and `clumpiness' of two-temperature coronae. 
}
\label{fig:multiphase}
\end{figure}

By forming a condensing inflow, conductive cooling allows plasma cooling out of the corona to feed the thin disc. Figure~\ref{fig:condensing_inflow} demonstrates this property of our ZNF solutions. We define the hot phase as plasma with a temperature $T \geq 5 T_0$\footnote{Our results are largely independent of our choice for the minimum temperature of the `hot phase,' at least for the runs with $\mathcal{A} = 10$. Using a minimum hot phase temperature of 3$T_0$ or 1.5$T_0$ results in a 60 per cent decrease and 50 per cent decrease in the hot phase mass, respectively, once conduction is turned on in the $\mathcal{A} = 10$ ZNF simulations. For $\mathcal{A} = 10^3$, turning on conduction does not affect the hot phase mass when the minimum hot phase temperature is 3$T_0$ or 1.5$T_0$.} and the mass in the corona as the total mass above $|z| = z_T$. Without conduction turned on, the hot phase comprises ${\approx}10$ to 25 per cent of the coronal plasma by mass for the $\mathcal{A} = 10$ runs and ${\approx}5$ per cent in the $\mathcal{A} = 10^3$ cases. Activating conduction at $t = 50$ orbits results in a sudden decrease in the hot phase mass as the corona cools via thermal conduction and a condensing inflow forms. The hot phase mass drops by a factor of ${\approx}3$ for weak cooling and ${\approx}2$ for strong cooling, and coronal plasma can rain onto the disc. This `coronal rain' carries very little mass---2 to 3 per cent of the disc's mass in the weakly cooled, $\mathcal{A} = 10$ simulations, and only 0.1~per cent of the disc's mass in the more strongly cooled runs. Thus, mass accretion rates should be unaffected by this condensing inflow. However, such an inflow may rain optically thick material into the optically thin surface layers of the disc, if this material can cool and condense during its infall (see \S\ref{sec:hard_to_soft}). 

\begin{figure*}
\hbox{
\includegraphics[width=1.0\textwidth]{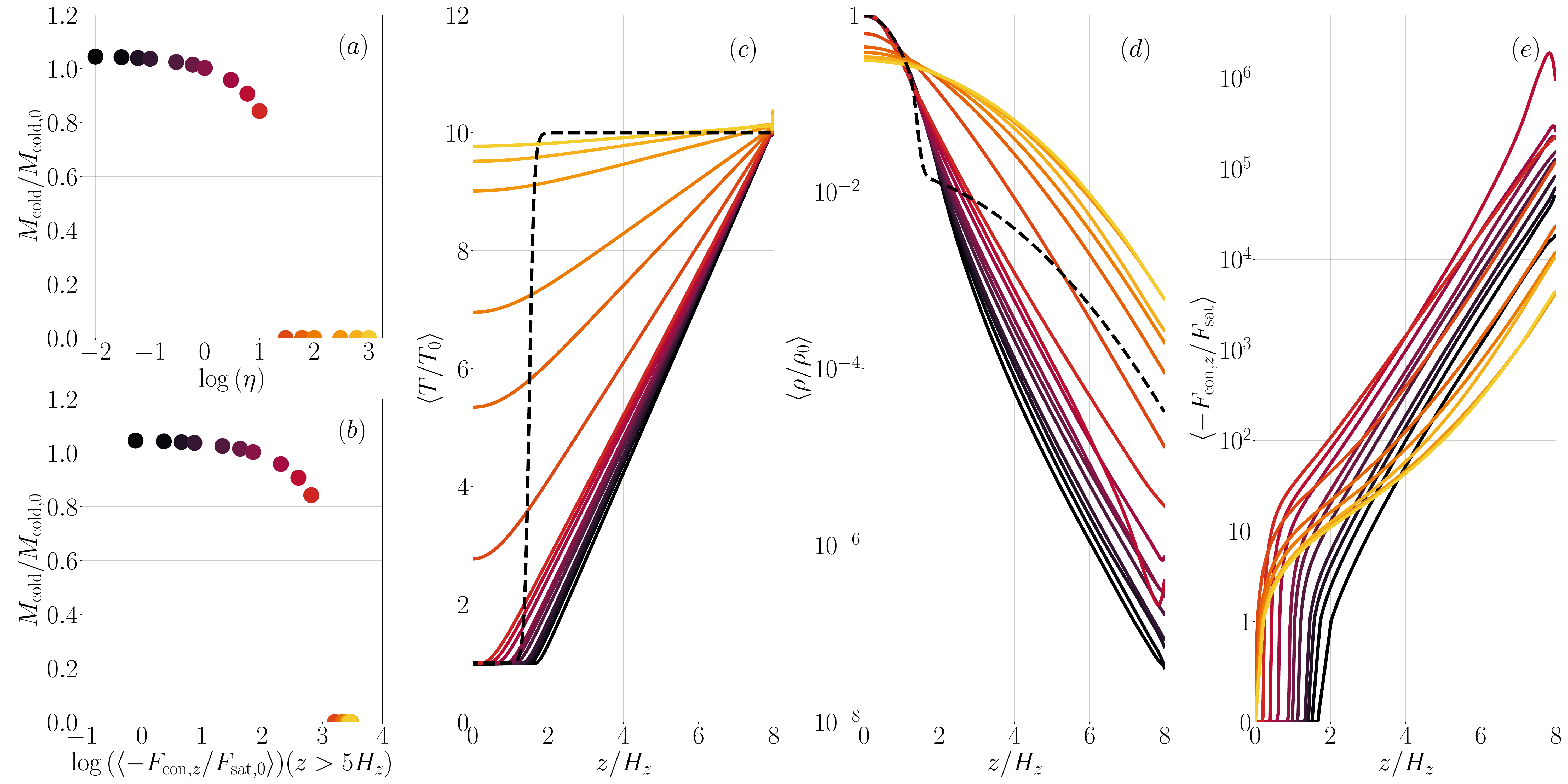}
}
\caption{Toy model results at time $\approx 89 H_z/c_{\rm s,0}$ into the simulation. The colours denote the value of the conduction amplification factor $\eta$ used in the toy model heat flux (see Equation~\ref{eq:F_eta}). (a) Cold plasma mass (defined as the mass of material with temperature $T \leq 1.5 T_0$) as a fraction of the initial cold plasma mass, $M_{\rm cold,0}$, as a function of $\eta$ and the conductive heat flux averaged above $z = 5 H_z$ (panel b). Temperature profiles (c) and density profiles (d) for all toy model simulations, with darker colours denoting lower values of $\eta$ and the black, dashed lines indicating the initial profiles. (e) Conductive heat flux profiles for all toy model simulations, measured in terms of the local saturated heat flux, $F_{\rm sat} \equiv 0.6 \rho c_{\rm s}^3$. We define the constant $F_{\rm sat,0} \equiv 0.6 \rho_0 c_{\rm s,0}^3$. Models do not evaporate, i.e., the cold plasma mass remains $M_{\rm cold} > 0$ for the duration of the simulation, unless $\eta > 10$, or equivalently, the heat flux into the disc is larger than $10^3 F_{\rm sat, 0}$. These heat fluxes are unphysically large, far above the free-streaming limit (see panel e). 
}
\label{fig:toy_model}
\end{figure*}

\subsection{Smoothing of multiphase temperature structure} \label{sec:multiphase}

Figure~\ref{fig:multiphase} shows log-density ($\rho$) and temperature ($T$) distributions measured at $|z|/H_z = 3$ for the ZNF $\beta_0 = 10$, $\mathcal{A} = 10$ simulations without conduction (i.e., from 30 to 50 orbits; top left) and with conduction active (from 70 to 100 orbits, long enough after conduction is turned on at 50 orbits for the solution to reach a steady state; bottom left). Beside these distributions, we show similar distributions for the NF, $\beta_0 = 10^4$, $\mathcal{A} = 10$ simulations without (top right) and with (bottom right)  conduction, measured at $|z|/H_z = 5$ from 30 to 100 orbits. As in \citetalias{Bambic2024}, we find that the solutions without conduction exhibit multiphase structure, with broad density and temperature distributions that form along curves of constant pressure (shown as light blue dotted curves in Fig.~\ref{fig:multiphase}). 

In the ZNF sims, this characteristic structure vanishes once conduction is activated at 50 orbits. Temperature distributions become sharply peaked, transitioning from a spread of ${\approx}11 T_0$ at $|z| = 3 H_z$ in the ZNF simulations to only ${\approx}3$--$4 T_0$ at the same height. The spreads in density remain virtually unchanged in both ZNF and NF simulations, although there may be a \textit{slight} broadening in log-density space in addition to a shift of the distributions with conduction toward densities lower than those characteristic of simulations without conduction. While a condensing inflow in the ZNF simulations with conduction results in a substantial drop in density, the NF simulations remain outflowing at all $|z| > 2 H_z$, and the mean density is unchanged by the inclusion of conduction.

\section{A toy model for evaporation} \label{sec:toy_model}

Conductive evaporation does not occur in any of our simulations. The geometric suppression of the heat flux plays some role in this outcome, decreasing the heat flux into the disc by an order unity factor. However, as we assessed in \S\ref{sec:timescales} and Figure~\ref{fig:timescales}, the dominant process opposing evaporation is the strong two-temperature cooling in the surface layers of the disc. This conclusion leads us to the simple question: `How large must the heat flux into the disc be to overcome two-temperature cooling?'

To answer this question, we design a toy model for the corona-disc system. Our goal is to set up an atmosphere in hydrostatic equilibrium (to control for the effects of the outflow) with a purely vertical magnetic field (to ensure that the heat flux is unsuppressed, with $f_s = 1$), where a hot corona sandwiches a cold disc (to ensure that the heat flux is toward the disc). We then turn on conduction and cooling from this initial set up, and observe whether or not an evaporative flow forms. Instead of using the free-streaming form of the heat flux \eqref{eq:F_free}, we use a Spitzer-esque heat flux of the form
\begin{equation} \label{eq:F_eta}
    \boldsymbol{F}_{\rm con} = - \eta \rho_0 c_{\rm s,0} H_z \: \grad T,
\end{equation}
where $\eta$ is a free parameter. Note that we intentionally use a constant thermal diffusivity set by the mid-plane density and temperature, which can lead to heat fluxes well in excess of the free-streaming value. By artificially varying $\eta$ and therefore the magnitude of the heat flux into the disc, we can assess how large the heat flux into the disc must be to form an evaporative flow. The choice of the heat flux is discussed further in \S\ref{sec:toy_set_up}. 

\subsection{Problem set-up} \label{sec:toy_set_up}

We restrict our calculation to one dimension, the $z$-direction, which extends from $-8 < z/H_z < 8$, although the results do not change for a domain with a maximum height of $6 H_z$. Such a tall domain minimizes the effect of the boundary condition on the thermal structure of the solutions near the mid-plane. The resolution chosen for this domain is the same as that used in the high-resolution three-dimensional simulations, 32 grid cells/$H_z$. Just as in the shearing-box simulations, we measure distances in terms of the initial mid-plane scale height $H_z$; however, these toy model simulations do not include the effects of rotation. Thus, we measure time in units of the free-fall time, $t_{\rm ff} \equiv H_z/c_{\rm s,0}$.  

Motivated by the peak temperature of ${\approx} 9 T_0$ achieved in the ZNF $\mathcal{A} = 10$ simulations without conduction, we initialize an atmosphere with a maximum temperature of $T_{\rm max} = 10 T_0$ above $|z| \approx 2 H_z$ (in the corona) and a minimum temperature of $T_0$ below this height (in the disc). We initialize the atmosphere with a temperature profile,
\begin{equation}
    T(z) = T_0 + \frac{9}{2} T_0 \Bigl\{ \tanh{\left[8 (|z| - 1.5) + 1 \right]} + 1 \Bigr\} .
\end{equation}
This profile is shown by the dashed black line in panel~(c) of Fig.~\ref{fig:toy_model}. The corresponding density profile is computed by solving the equation of hydrostatic equilibrium for the disc's local gravitational potential, $\Phi = (c_{\rm s,0}/H_z)^2 z^2$. As a result, we achieve a similar initial density profile within the disc to that exhibited in the steady state of the ZNF and weak NF simulations within the same region of $|z| < 2 H_z$ (Fig.~\ref{fig:temperature_density}). Above $|z| = 2 H_z$, the density in the toy-model initial condition drops off much less steeply compared to the density profiles in the three-dimensional simulations. Because our atmosphere is initially hydrostatic and not outflowing, the initial density in the corona is higher than that of the coronae of the ZNF and weak NF simulations.

The upper and lower boundaries are handled in the same way as was done for the fiducial simulations, i.e., the conductive heat flux $\boldsymbol{F}_{\rm con}$ is copied into ghost zones with zero gradient. Unlike the fiducial simulations, however, we pin the outer boundary temperature to $10 T_0$. This condition ensures that there is a continuous heat flux from the outer boundary into the domain and toward the disc. We choose a high value of $v_{\rm max} = 10^3 \: H_z \Omega$ to ensure that, even with artificially high heat fluxes well in excess of the free-streaming limit, the solutions remain accurate. The heat flux into the disc is removed through our cooling function \eqref{eq:cooling_function}, where we choose $\mathcal{A} = 10$ to match our most weakly cooled three-dimensional models. We choose an extremely low floor of $\rho_{\rm floor}/\rho_0 = 10^{-10}$, although the solutions never come within an order of magnitude of this floor density. 

While we tried to implement the free-streaming heat flux in these toy model simulations, we found that this form of the heat flux does not achieve a steady state sufficient to allow for a measurement of evaporation. The atmosphere collapses onto the disc, causing a shock wave, which reverses the direction of the heat flux such that conduction channels heat out of the disc. The reason for this behaviour is that, unlike the heat flux used in our toy model (Equation~\ref{eq:F_eta}), the free-streaming heat flux depends on density $\rho$. Because of the large density gradient (relative to the temperature gradient), in the corona-disc transition zone, the heat flux nearest the disc is much higher than that higher up in the corona. As a result, the bottom of the corona cools first and begins to collapse onto the disc before sufficient thermal energy from higher in the corona can arrive to offset this cooling. Thus, no steady state with the free-streaming heat flux is possible, and using the free-streaming heat flux would prevent us from accurately measuring the magnitude of the heat flux required to induce evaporation. 

We run the simulation until $t \approx 89 t_{\rm ff}$. By this time, the solutions have relaxed into a steady state, and the fate of the disc, whether the disc evaporates or remains intact, is sealed for all time.

\subsection{Toy model results} \label{sec:toy_results}

Figure~\ref{fig:toy_model} shows the results of these calculations for 16 values of $\eta$ in the range $10^{-2} \leq \eta \leq 10^3$. We measure the cold gas mass $M_{\rm cold}$, defined as the mass of material with temperature $T \leq 1.5 T_0$, relative to the initial cold gas mass $M_{\rm cold,0}$ measured at $t = 0$. As is clear from panels~(a) and (b) of Fig.~\ref{fig:toy_model}, the cold disc is completely destroyed by $t \approx 89 t_{\rm ff}$ for any heat flux corresponding to $\eta > 10$. 

Panel~(b) of Fig.~\ref{fig:toy_model} shows the final cold gas mass at $t \approx 89 t_{\rm ff}$ as a function of the heat flux $F_{\rm con}$ averaged above $z > 5 H_z$. We measure the heat flux relative to the peak saturated value achievable for a plasma with temperature $T_0$ and density $\rho_0$, i.e., $F_{\rm sat, 0} \equiv 0.6 \rho_0 c_{\rm s, 0}^3$. Profiles of heat flux relative to the local saturated value, $F_{\rm sat} \equiv 0.6 \rho c_{\rm s}^3$, are shown in panel~(f). The critical heat flux for evaporation is $F_{\rm crit} \approx 10^3 \: F_{\rm sat, 0}$---far larger than any heat flux that can be physically attained in the system. Similarly, models that evaporate display heat fluxes of $|F_{{\rm con},z}|/F_{\rm sat} \gtrsim 10$ above $|z| = 2 H_z$. Two-temperature cooling is so strong in the disc's surface layers that only a completely unphysical heat flux can induce evaporation. This analysis underscores the primary conclusion of this paper: ion heat fluxes from the corona into the disc are insufficient to evaporate the disc, even when the heat flux achieves its maximum value at the free-streaming limit.

\section{Discussion} \label{sec:discussion}

\subsection{Comparison to previous work} \label{sec:previous_work}

Our conclusions are similar to \citetalias{Cho2022}, in that we find that conduction in the inner regions of XRB accretion discs is more likely to lead to a condensing inflow from the corona onto the disc than to result in disc evaporation. Yet, we arrive at our conclusions by examining very different physical mechanisms: free-streaming ion thermal conduction vs. Spitzer electron conduction, and two-temperature ion cooling via Coulomb collisions rather than Bremsstrahlung cooling of a one-temperature plasma. 

\subsection{Limitations of our model} \label{sec:limitations}

Physical effects not captured in our simple two-temperature model may be important for understanding the role of conduction in disc evaporation. In the bulk of the disc where the plasma becomes one-temperature, electron thermal conduction dominates the heat flux. Electron conduction is likely to operate in the Spitzer regime (as in \citetalias{Cho2022}), rather than in the free-streaming regime. Further, radiation transport may substantially modify the temperature structure of the disc's surface layers, potentially forming a steep temperature gradient approaching the mid-plane. 

In addition, the form of our two-temperature cooling function \eqref{eq:cooling_function} implies that cooling becomes stronger with increasing temperature. Bremsstrahlung cooling behaves similarly, with the cooling rate scaling ${\propto}T_e^{1/2}$. However, if the disc is radiation pressure dominated, gas temperatures may be low enough that substantial line cooling takes place near the mid-plane. At $T \approx 10^{5-6} \: {\rm K}$, the cooling rate \textit{decreases} with increasing temperature, and conductive heating would lead to even weaker cooling. Such thermally unstable heating may play some role in the bulk of the disc. Indeed, this is the regime relevant for the solar corona and for white dwarf coronae, where conductive evaporation is traditionally studied. We note that such low temperatures are difficult to achieve in XRB discs, whose inner edges are hot enough to emit soft, thermal X-rays. The $10^6 \: {\rm K}$ branch of the cooling curve is more likely to be approached at larger radii from the black hole in XRB discs, or in much cooler, much more radiation dominated flows, such as those fueling AGN. 

\subsection{Soft-to-hard state transitions in XRBs} \label{sec:soft_to_hard}

Our primary motivation for exploring ion thermal conduction was to find the long-sought physical mechanism that causes disc evaporation, truncation, and therefore, state transitions in XRB accretion flows. Based on our calculations, ion thermal conduction cannot serve this role. Heat fluxes orders of magnitude larger than what can be mustered by free-streaming ion conduction are necessary to force evaporation in the presence of two-temperature cooling.

The presence and magnitude of NF, rather than the inclusion of thermal conduction, has the most substantial effect on our two-temperature disc-corona systems. Simulations with zero net vertical magnetic flux (ZNF) do not evaporate, but instead, their coronae cool via conduction and form condensing inflows onto the disc. Similarly, models with NF launch magnetocentrifugal outflows. Such outflows may act to deplete the disc, reducing its surface density until Coulomb collisions become sufficiently infrequent that the flow becomes two-temperature. The runaway ion heating that ensues might evaporate the disc. Similarly, magnetic pressure support afforded by NF fields allows for lower disc densities and optical depths compared to those in gas-pressure-dominated and radiation-pressure-dominated flows \citep{Mishra2022, Huang2023}. These low densities, combined with inefficient cooling at these reduced optical depths, encourage runaway ion heating and evaporation. Thus, the introduction of NF, through providing magnetic pressure support and through launching winds that deplete the disc, might be the agent of disc evaporation and truncation in XRBs \citep{Ferreira1997,Ferreira2006,Begelman2014,Liska2023}. 

\subsection{Hard-to-soft transitions and iron lines} \label{sec:hard_to_soft}

Conduction acts as an efficient cooling mechanism in coronae threaded by ZNF, producing a condensing inflow onto the disc. In this way, just as the introduction of NF from larger scales may prompt the evaporation of the disc through wind depletion, the removal of NF and the subsequent cessation of such a wind may lead to sudden condensation of the inner hot RIAF into a disc. Indeed, our simulations show that the introduction of conduction into the two-temperature ZNF simulations can decrease the mass of plasma in the hot phase of the corona by as much as 80 per cent in the weakly cooled, $\mathcal{A} = 10$ simulations (Fig.~\ref{fig:condensing_inflow}). If this physics of condensation of the corona onto the disc operates in real systems, it could contribute to a hard-to-soft state transition and the re-formation of Fe lines in the inner regions of XRBs. Thus, the impact of ion thermal conduction on the disc-corona system is intimately tied to whether or not NF fields thread the accretion flow. 

While the presence and strength of the Fe line signal may be affected by conductive cooling and condensation in the corona, the hard X-ray power-law emission produced by coronae is likely unaffected by conduction. Even though conduction significantly decreases the spread in temperature in the coronae of our models (Fig.~\ref{fig:multiphase}), the spread in density is virtually unchanged. Since we are evolving the ion temperature rather than the lepton temperature, the temperature spread would be unlikely to affect the cut-off in the X-ray spectrum. Similarly, the negligible effect on the density spread implies that the optical depth to electron scattering, and therefore the power-law index measured from observations, is unlikely to be changed by free-streaming ion conduction. 

\section{Summary and conclusion} \label{sec:conclusion}

We have implemented field-aligned thermal conduction at the saturated, free-streaming limit, into local simulations of vertically stratified accretion flows. By treating the ions in the corona as a single MHD fluid subject to cooling via Coulomb collisions with radiatively efficient electrons, we capture temperature inversions reminiscent of a hot corona surrounding a colder disc, which then direct a conductive heat flux from the corona into the disc. Through the application of our two-temperature model with conduction to a large suite of three-dimensional stratified shearing-box simulations and a one-dimensional, highly idealized, toy model, we have shown that ion thermal conduction is unable to evaporate thin accretion discs into RIAFs, as has previously been proposed \citep{Spruit2002}. Instead, two-temperature cooling in the surface layers of discs removes energy from the ions, radiating this thermal energy away before any substantial conductive heating can impact the bulk of the dense disc.

Our main results are as follows:
\begin{enumerate}
    \item All of our two-temperature models with thermal conduction form temperature inversions (Figure~\ref{fig:temperature_density}, left panels), with a hotter corona surrounding a colder disc. In NF simulations, conduction has no effect on the ion temperatures in the corona. For ZNF simulations, conduction acts to cool coronae, decreasing their maximum horizontally and temporally averaged temperatures by about 25 per cent in the weakly cooled, $\mathcal{A}  = 10$ simulations, independent of the initial strength of the ZNF fields.
    \item The temperature inversions allow for a net conductive heat flux from the hot corona into the cold disc. This heat flux is not negligible. In discs thinner than $H_z/R_0 \approx$ a few $\times 10^{-2}$, conductive evaporation would destroy the discs more rapidly than inflow could empty them, \textit{if cooling were ignored}.
    \item Cooling through Coulomb collisions between hot ions and rapidly cooling leptons far outpaces the heating delivered by thermal conduction into the disc's surface layers. The two-temperature cooling timescale is shorter than the conductive heating timescale throughout the coronae of all of our simulations (\S4.3 and Fig.~\ref{fig:timescales}), independent of the presence and strength of NF and the strength of Coulomb cooling (as parameterized through the Coulomb coupling parameter, $\mathcal{A}$) in the corona. 
    \item The dominantly toroidal magnetic fields formed through MRI turbulence and Keplerian shear in the stratified shearing box act to geometrically suppress the heat flux from the corona into the disc. We find suppression factors of $\langle f_s \rangle_{r \varphi t} \approx 0.1-0.6$ (Fig.~\ref{fig:fs_profiles}). The heat flux is less suppressed in moderate NF simulations, which exhibit field lines that are more rigid and more vertically aligned, on average compared to ZNF simulations. Higher-resolution simulations show less geometric suppression of the heat flux, and estimates based on post-processing generally under-estimate the heat flux into the disc.
    \item Using a one-dimensional toy model of the disc-corona system, where the `corona' above two mid-plane scale heights is $10 \times$ hotter than the disc and the system is initially in hydrostatic equilibrium, we showed that the heat flux from the corona into the disc must be unphysically large to overcome two-temperature cooling in the disc and induce evaporation (\S\ref{sec:toy_model} and Fig.~\ref{fig:toy_model}). 
    \item Rather than forming outflows, ZNF simulations with conduction form condensing inflows, which can remove as much as 80 per cent of the hot phase plasma mass from the coronae (\S4.5 and Fig.~\ref{fig:condensing_inflow}). However, because of how little mass is contained in the corona, this condensing inflow is unlikely to affect the accretion rate or luminosity of the system.
    \item Thermal conduction smooths out multiphase temperature structure, shrinking the spread in temperature distributions and eliminating the characteristic nearly isobaric density-temperature structure at fixed height $|z|$ that was observed in the two-temperature simulations of Paper I. The spread in densities, and consequently, observational signatures that can be probed by the power-law index of the X-rays released by the corona, are unaffected by the inclusion of thermal conduction.
\end{enumerate}

Our work indicates that ion thermal conduction cannot provide the physical mechanism that enables soft-to-hard state transitions in XRBs. The introduction of NF changes the vertical structure of two-temperature coronae from a condensing inflow cooled via thermal conduction into a magnetocentrifugal outflow. Similarly, removal of NF results in condensation of a hot corona and the feeding of a thin, optically thick accretion disc near the ISCO. The presence of significant magnetic support also decreases disc densities, leading to less efficient cooling and promoting the formation of a RIAF (e.g., \citealt{Mishra2022} and \citealt{Huang2023}). We thus speculate that a change in the amount of NF threading an accretion flow, either through in-situ generation of poloidal flux via a dynamo or transport of this flux from larger scales, is responsible for the state transitions observed ubiquitously in XRBs, consistent with works by \cite{Begelman2014} and recently, \cite{Liska2023}. 

\section*{Acknowledgements}

CJB would like to thank Mitch Begelman and Omer Blaes for stimulating discussions that improved this work. CJB is supported by the National Science Foundation (NSF) Graduate Research Fellowship.  This work was supported in part by a Simons Investigator award to EQ from the Simons Foundation. Computational resources for our simulations were provided by the NSF's ACCESS program (formerly XSEDE) under grants PHY220078 and PHY230170 on Purdue's ANVIL supercomputer, and through the Flatiron Institute's Center for Computational Astrophysics (CCA). Analysis was performed on the Stellar supercomputer, operated through the Princeton Institute for Computational Science and Engineering (PICSciE) and the Office of Information Technology's High Performance Computing Center at Princeton University.

\section*{Software}

The MHD simulations presented in this work were performed using the \textit{Athena++} code \citep{Stone2020}. Three-dimensional renderings were produced using the VisIt software package \citep{Childs_VisIt_An_End-User_2012}, which is supported by the Department of Energy with funding from the Advanced Simulation and Computing Program, the Scientific Discovery through Advanced Computing Program, and the Exascale Computing Project. Analysis was performed using \textit{numpy} \citep{numpy}, figures were produced through \textit{matplotlib} \citep{matplotlib}, and colour maps/schemes were imported from the \textit{cmasher} package \citep{cmasher}.

\section*{Data availability}

Simulation data is available upon reasonable request to the corresponding author.


\begin{appendix}

\section{Sound wave tests} \label{sec:sound_waves}

To test the robustness of our code, here we demonstrate that our chosen two-moment method is able to capture the correct dispersion and damping of linear sound waves in the presence of thermal conduction. We begin by perturbing the MHD equations for mass, momentum, and energy conservation around a static background with uniform density $\rho_0$, pressure $P_0$, temperature $T_0$, and magnetic field $\boldsymbol{B} = B_0 \hat{x}$, where we work in one dimension ($x$). For this analysis, we adopt a Spitzer form for the heat flux,
\begin{equation} \label{eq:Spitzer_appendix}
    \boldsymbol{F}_{\rm con} = - \kappa_{\parallel} \eb \eb \bcdot \grad  T ,
\end{equation}
where $\kappa_{\parallel}$ is the thermal conductivity along field lines. Standard linear theory for a perturbation with complex frequency $\omega$ and real wave-vector $\boldsymbol{k} = k_{\parallel} \eb$ yields the following relation between density perturbations $\delta \rho$ and pressure perturbations $\delta P$,
\begin{equation} \label{eq:p_rho}
    \frac{\delta P}{P_0} = \left( \frac{-i \omega \frac{\gamma}{\gamma - 1} + k_{\parallel}^2\chi}{-i \omega \frac{1}{\gamma - 1} + k_{\parallel}^2\chi} \right) \frac{\delta \rho}{\rho_0} \equiv a^2 \frac{\delta\rho}{\rho_0}.
\end{equation}
Here, we have introduced the quantity $\chi \equiv \kappa_{\parallel} T_0/P_0$. Because the perturbed continuity and momentum equations yield a dispersion relation $\omega^2 = a^2 c_{\rm{s},0}^2 k_{\parallel}^2$, we can write the full dispersion relation as
\begin{equation} \label{eq:dispersion_relation}
    \omega^2 \left( -\imag \omega \frac{1}{\gamma - 1} + \omega_{\rm c} \right) = k_{\parallel}^2 c_{\rm s, 0}^2  \left( -\imag \omega \frac{\gamma}{\gamma - 1} + \omega_{\rm c} \right),
\end{equation}
where we have introduced the conduction frequency, $\omega_{\rm c} \equiv k_{\parallel}^2\chi$.

\begin{figure}
\hbox{
\includegraphics[width=0.48\textwidth]{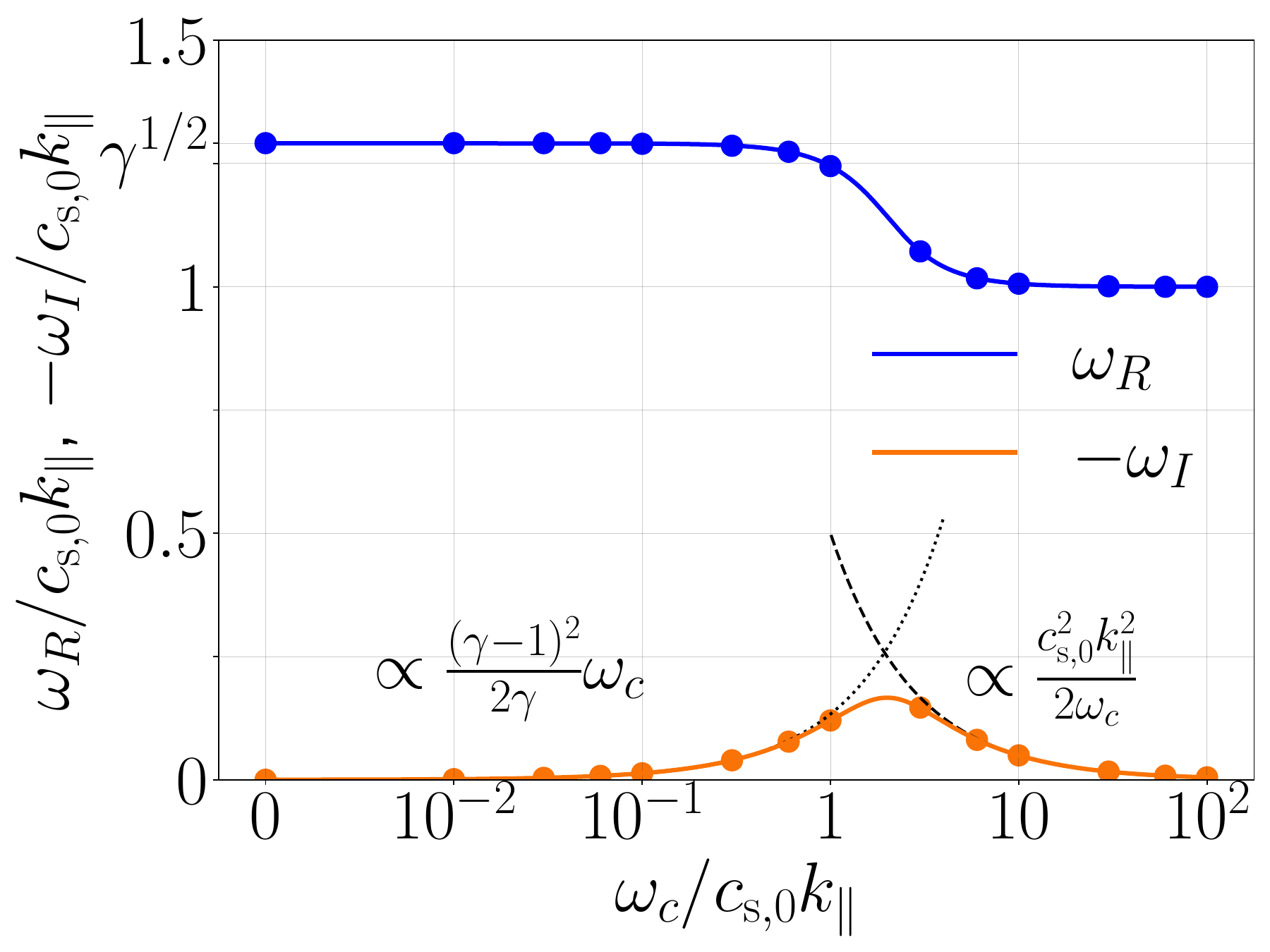}
}
\caption{Dispersion relation for a sound wave subject to thermal conduction in the Spitzer regime \eqref{eq:Spitzer_appendix}. Solid lines show the numerical solution to the dispersion relation \eqref{eq:dispersion_relation}, while points indicate measurements based on a series of linear sound wave tests (see text). The real frequency $\omega_R$ is indicated with the color blue, while the imaginary frequency/ damping rate $-\omega_I$ (the minus sign indicates damping of waves) is indicated by the color orange.  
}
\label{fig:dispersion_relation}
\end{figure}

The solid lines in Figure~\ref{fig:dispersion_relation} show the real (blue) and imaginary (orange) parts of $\omega$ as a function of $\omega_{\rm c}$. To gain an intuition for the form of the curves $\omega (\omega_{\rm c})$ in the limits of slow conduction ($\omega \gg \omega_{\rm c}$) and fast conduction ($\omega \ll \omega_{\rm c}$), we expand $\omega$ in powers of a small parameter $\varepsilon$ such that $\omega = \omega_0 + \varepsilon \omega_1 + \dots$. A natural choice for $\varepsilon$ is $\varepsilon_{\rm s} = \omega_{\rm c}/\omega$ for `slow' conduction and $\varepsilon_{\rm f}= \omega/\omega_{\rm c}$ for `fast' conduction. Inserting this form for $\omega$ into the dispersion relation \eqref{eq:dispersion_relation}, we analyze the resulting equations order by order.

In the limit of slow conduction ($\omega \gg \omega_{\rm c}$), the $\mathcal{O}(1)$ and $\mathcal{O}(\varepsilon_{\rm s})$ equations imply
\begin{equation}
    \omega = \underbrace{k_\parallel \gamma^{1/2} c_{\rm s,0} }_{\omega_0} \:\underbrace{- \imag \frac{(\gamma - 1)^2}{2 \gamma} \omega_{\rm c}}_{\varepsilon_{\rm s} \omega_1}.
\end{equation}
Thus, when conduction is slow, sound waves behave like adiabatic sound waves with sound speed $\gamma^{1/2} c_{\rm s, 0}$, but with the added effect of weak damping. When conduction is fast ($\omega \ll \omega_{\rm c}$), the $\mathcal{O}(1/\varepsilon_{\rm f})$ and $\mathcal{O}(1)$ equations imply
\begin{equation}
    \omega = \underbrace{k_\parallel c_{\rm s,0} }_{\omega_0} \:\underbrace{- \imag \frac{k_{\parallel}^2c_{\rm s, 0}^2 }{2 \omega_{\rm c}}}_{\varepsilon_{\rm f} \omega_1}.
\end{equation}
These are nearly isothermal sound waves, whose departure from isothermality is damped by conduction. Thus, sound waves propagate at different speeds depending on the strength of thermal conduction, and they damp at rates controlled by $\omega_{\rm c}$.

We perform a series of simulations of a one-dimensional sound wave in a periodic box of size $L_x$. The background density $\rho_0 \equiv 1$ and pressure $P_0 \equiv 1$ are defined to ensure the isothermal sound speed $c_{\rm s, 0} \equiv 1$. We initialize a pure eigenmode of the perturbed fluid equations, such that the initial density is given by
\begin{equation}
    \rho(x) = \rho_0 + \xi \rho_0 \cos{\left( k_{\parallel} x \right)},
\end{equation}
where $k_{\parallel}~=~2 \pi/ L_x$ and we choose $\xi = 10^{-6}$ to ensure that the waves are well described by linear theory. To sample the dispersion relation, we choose 14 values of the conduction frequency $\omega_{\rm c}$ ranging from 0 to $10^2k_{\parallel} c_{\rm s, 0}$. We choose $v_{\rm max} = 30 \sqrt{2} c_{\rm s,0}$ to match the same ratio of $v_{\rm max}/c_{\rm s,0}$ that was used in the three-dimensional shearing-box simulations, $v_{\rm max} = 30 H_z \Omega$. Sound waves are run for 10 box-crossing times $t_{\rm cross} \equiv L_x/c_{\rm s,0}$, and we compute the real frequencies and damping rates of the waves from the time-series of $\rho$ measured at $x = L_x/2$. The results are shown in Figure~\ref{fig:dispersion_relation}.

The close agreement between linear theory and our measured sound wave frequencies and damping rates indicate that the code handles conduction accurately. Turning down $v_{\rm max}$ to $3 \sqrt{2} c_{\rm s,0}$ has only a minor effect on the agreement between the numerical and physical dispersion relations, indicating that the code is quite robust. Choices of $v_{\rm max} \lesssim c_{\rm s, 0}$ result in the completely wrong dispersion relation, with real frequencies pinned at $\omega_{\rm R} = \gamma^{1/2} c_{{\rm s},0} k_{\parallel}$, independent of $\omega_{\rm c}$, and no damping of the waves. 

\section{Effect of Reduced Speed of Light} \label{sec:vmax_appendix}

\begin{figure*}
\hbox{
\includegraphics[width=1.0\textwidth]{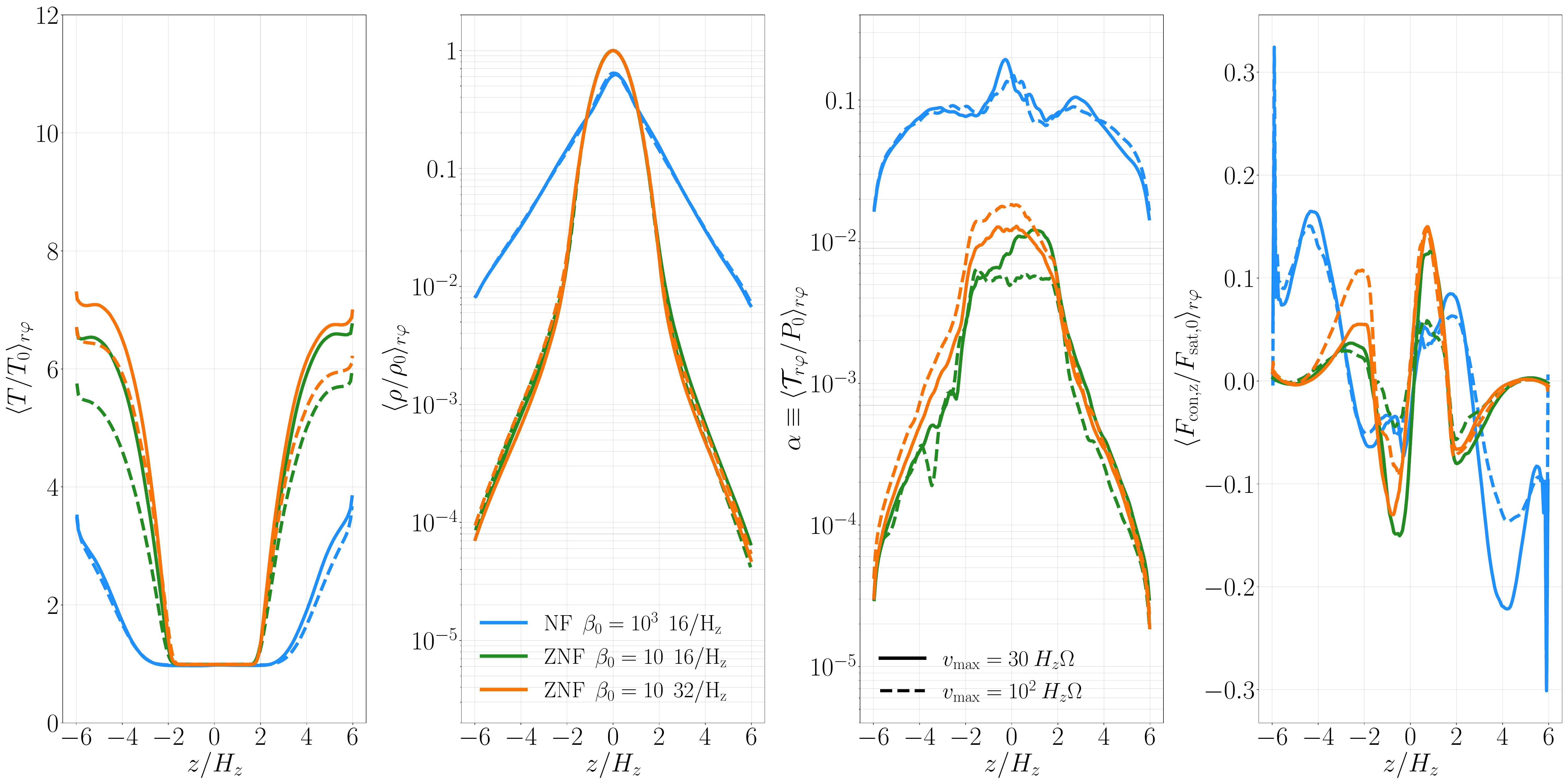}
}
\caption{Horizontally and temporally averaged temperature $T$, density $\rho$, turbulent stress $\alpha$, and vertical heat flux $F_{{\rm con}, z}$ profiles for $\mathcal{A} = 10$ ZNF and NF simulations with different $v_{\rm max}$ at $t = 50$ orbits. Colour denotes the simulation parameters (e.g. ZNF vs. NF, $\beta_0$, etc.) while the linestyle corresponds to $v_{\rm max}$, with $v_{\rm max} = 30 \: H_z \Omega$ denoted by solid lines and $v_{\rm max} = 10^2 \: H_z \Omega$ denote by dashed lines. Time averaging of the profiles occurs over the intervals $(t_i, t_f) = (30, 100)$ orbits for the NF $v_{\rm max} = 30 \: H_z \Omega$ run and $(t_i, t_f) = (50, 100)$ orbits for the NF $v_{\rm max} = 10^2 \: H_z \Omega$ run. For the high resolution ZNF simulation, both runs with different $v_{\rm max}$ are averaged from $(t_i, t_f) = (70, 100)$ orbits, as is the fiducial resolution ZNF simulation with $v_{\rm max} = 30 \: H_z \Omega$. 
The fiducial resolution ZNF $v_{\rm max} = 10^2 \: H_z \Omega$ run is averaged over the interval $(t_i, t_f) = (85, 100)$ orbits. The profiles are similar, independent of $v_{\rm max}$.
}
\label{fig:vmax_profiles}
\end{figure*}

The two-moment method accurately returns a heat flux comparable to the saturated value so long as Equation~\ref{eq:vmax_condition} is satisfied. Here, we assess if our chosen $v_{\rm max}$ is sufficient to yield accurate results in our simulations with thermal conduction. We restart three simulations at $t = 50$ orbits with $v_{\rm max}$ increased to $v_{\rm max} = 10^2 \: H_z \Omega$ and run these simulations until $t = 100$ orbits. These three simulations are the ZNF $\mathcal{A} = 10$ $\beta_0 = 10$ model at 16 cells/$H_z$ resolution, the ZNF $\mathcal{A} = 10$ $\beta_0 = 10$ model at 32 cells/$H_z$ resolution, and the NF $\mathcal{A} = 10$ $\beta_0 = 10^3$ simulation at 16 cells/$H_z$ resolution. We choose these particular simulations because we find that the ZNF simulations are more likely to exhibit cells where the true heat flux evolved in the simulation is less than the free-streaming value, i.e., $|\boldsymbol{F}_{\rm con}|/F_{\rm sat} < 1$, where $F_{\rm sat} \equiv 0.6 \rho c_s^3$. In these zones, $v_{\rm max}$ is too small to satisfy Equation~\ref{eq:vmax_condition}. Similarly, the moderate NF simulations, with their larger velocities, are more likely to introduce advective errors that increase the size of the time-derivative term in  Equation~\ref{eq:second_moment}. 

Figure~\ref{fig:vmax_profiles} shows profiles of the horizontally averaged temperature, density, turbulent $\alpha$ parameter, and vertical conductive heat flux for these three simulations. The solid lines denote the results computed with $v_{\rm max} = 30 \: H_z \Omega$ (the fiducial $v_{\rm max}$ used throughout the paper) while dashed lines show the results after increasing $v_{\rm max}$ to $10^2 \: H_z \Omega$ at 50 orbits. The temperature, density, $\alpha$ and vertical heat flux profiles are relatively similar between simulation runs presented in the main paper and runs with increased $v_{\rm max}$, indicating that our chosen $v_{\rm max}$ is sufficiently large that the heat flux is reasonably close to the saturated value.
Just as in the profiles with $v_{\rm max} = 30 \: H_z \Omega$, increasing $v_{\rm max}$ does not result in evaporation of the cold disc, in agreement with the conclusions presented throughout the main paper.

\begin{figure}
\hbox{
\includegraphics[width=0.5\textwidth]{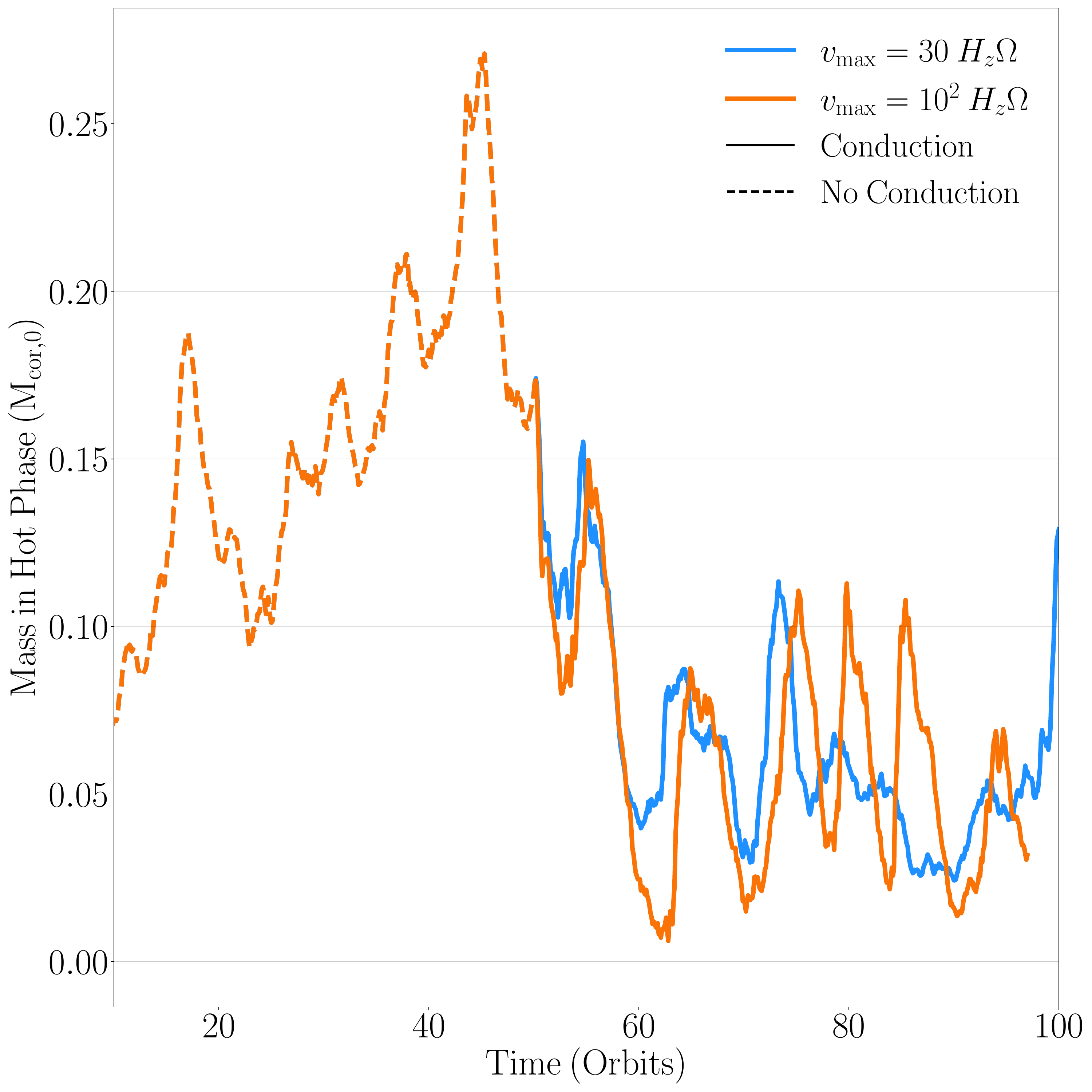}
}
\caption{Evolution of hot phase $(T > 5 T_0)$ in the corona as a function of time for the high resolution ZNF $\mathcal{A} = 10$ $\beta_0 = 10$ simulation, similar to Figure~\ref{fig:condensing_inflow}, for different values of $v_{\rm max}$. Restarting the simulation at $t = 50$ orbits with a higher value of $v_{\rm max}$ has a negligible effect on the time-averaged amount of hot gas in the corona (and thus the amount of material rained onto the disc). The solutions diverge from one another over time due to chaos intrinsic to the turbulent system. 
}
\label{fig:vmax_condensation}
\end{figure}

Similarly, in Figure~\ref{fig:vmax_condensation}, we demonstrate that our conclusion that thermal conduction causes a condensing inflow onto the disc in simulations without NF (see Fig.~\ref{fig:condensing_inflow}) does not depend on the chosen value of $v_{\rm max}$. Finally, we note that if $v_{\rm max}$ is not large enough, such that Equation~\ref{eq:vmax_condition} is not satisfied, the two-moment method is more likely to under-estimate the total heat flux rather than over-estimate this flux. Thus, we checked that the profiles of the horizontally averaged conductive heat flux produced by the simulations, i.e. $\langle F_{{\rm con},z} \rangle_{r \varphi}$ were in agreement with the expectations from the free-streaming value: $\langle F_{\rm sat} (F_{{\rm con},z}/|F_{\rm con}|) \rangle_{r \varphi}$. To compute these profiles, we restricted our computations to hot zones in the corona where $T > 1.5 T_0$, since in the disc midplane, a uniform temperature would imply no heat flux while $F_{\rm sat}$ would be $0.6 \rho_0 c_{s,0}^3$. We find that the post-processed and simulated heat fluxes agree to within 5 per cent error for all simulations, with smaller errors corresponding to higher resolution and higher $v_{\rm max}$ runs. 

The constraint on $v_{\rm max}$ implied by Equation~\ref{eq:vmax_condition} is more difficult to satisfy if $|\eb\bcdot\grad \ln T|$ is small.  The reason is that the conductivity $\kappa$ in Equation~\ref{eq:kappa} is then larger and so a larger $v_{\rm max}$ is necessary for the heat flux to equal the equilibrium (saturated) value in the two-moment method. The dominantly toroidal nature of the field lines in our local disc patches implies that temperature will be relatively uniform horizontally (e.g., the narrow temperature distributions at fixed height in the ZNF simulations in Fig.~\ref{fig:multiphase}).  Thus with our chosen values of $v_{\rm max}$ we somewhat underestimate the horizontal heat fluxes relative to the correct saturated value. However, when field lines possess a significant vertical component, the field-aligned temperature gradient samples a large variation in temperature, $\kappa$ is small enough to be controlled by the chosen $v_{\rm max}$, and the heat fluxes are accurate such that $|F_{\rm con}| \approx F_{\rm sat}$. These zones with substantial vertical components to the field lines dominate the vertical transport of heat and thus mediate conductive heating, conductive cooling, and evaporation (or lack thereof) in our simulations.

\end{appendix}

\end{document}